\documentclass[%
reprint,
superscriptaddress,
shortbibliography,
 amsmath,amssymb,
 aps,
]{revtex4-1}

\usepackage{graphicx}
\usepackage{dcolumn}
\usepackage{bm}
\usepackage{tabularx}
\usepackage{hyperref}
\begin{document}

\title{Supercurrent in ferromagnetic Josephson junctions with heavy metal interlayers. II. Canted magnetization}

\author{Nathan~Satchell}
\affiliation{Department of Physics and Astronomy, Michigan State University, East Lansing, MI 48912, USA}
\affiliation{School of Physics and Astronomy, University of Leeds, Leeds, LS2 9JT, United Kingdom}

\author{Reza~Loloee}
\affiliation{Department of Physics and Astronomy, Michigan State University, East Lansing, MI 48912, USA}

\author{Norman~O.~Birge}
\email{birge@pa.msu.edu}
\affiliation{Department of Physics and Astronomy, Michigan State University, East Lansing, MI 48912, USA}

\date{\today}

\begin{abstract}

It has been suggested by theoretical works that equal-spin triplet pair correlations can be generated in Josephson junctions containing both a ferromagnet and a source of spin-orbit coupling. Our recent experimental work suggested that such triplet correlations were not generated by a Pt spin-orbit coupling layer when the ferromagnetic weak link had entirely in-plane anisotropy (N. Satchell and N.O. Birge, Phys. Rev. B \textbf{97}, 214509 (2018)). Here, we revisit the experiment using Pt again as a source for spin-orbit coupling and a [Co(0.4 nm)/Ni(0.4 nm)]$_{\times8}$/Co(0.4 nm) ferromagnetic weak link with both in-plane and out-of-plane magnetization components (canted magnetization). The canted magnetization more closely matches theoretical predictions than our previous experimental work. Our results suggest that there is no supercurrent contribution in our junctions from equal-spin triplet pair correlations. In addition, this work includes the first systematic study of supercurrent dependence on Cu interlayer thickness, a common additional layer used to buffer the growth of the ferromagnet and which for Co may significantly improve the growth morphology. We report that the supercurrent in the [Co(0.4 nm)/Ni(0.4 nm)]$_{\times8}$/Co(0.4 nm) ferromagnetic weak links can be enhanced by over two orders of magnitude by tuning the Cu interlayer thickness. This result has important application in superconducting spintronics, where large critical currents are desirable for devices.

\end{abstract}


\maketitle

\section{Introduction}

Artificial superconducting - ferromagnetic (\textit{S}-\textit{F}) hybrids have been an area of intense research effort for over a decade due to the new physics which arises when superconducting pair correlations traverse the exchange field of a ferromagnet \cite{RevModPhys.77.935,RevModPhys.77.1321,eschrig_spin-polarized_2011, linder_superconducting_2015, 0034-4885-78-10-104501,doi:10.1098/rsta.2015.0150}. Of particular interest is the study of equal-spin triplet pair correlations (the $m_{\text{s}} = \pm1$ triplet components), which in Josephson junction experiments penetrate further into the \textit{F} layer than the spin-singlet and $m_{\text{s}} = 0$ triplet component, due to the two electrons propagating in the same spin band \cite{PhysRevLett.86.4096,keizer2006spin, PhysRevLett.96.157002,PhysRevLett.104.137002, robinson2010controlled, PhysRevB.82.100501}. Experimentally, equal-spin triplets are generated reliably in \textit{S--F'--F--F''--S} Josephson junctions, where the \textit{F', F''} ``spin-mixer layers'' mediate the conversion of singlet to triplet pair correlations \cite{PhysRevB.76.060504}.

Several theoretical studies have shown that spin-orbit coupling can act to convert spin-singlet correlations into equal-spin triplet correlations, potentially removing the need for ferromagnetic spin-mixer layers in Josephson junctions \cite{doi:10.1063/1.4743001, PhysRevLett.110.117003, PhysRevB.89.134517, Konschelle2014, PhysRevB.92.024510, alidoust2015spontaneous, alidoust2015long, jacobsen2016controlling, PhysRevB.95.024514, doi:10.7566/JPSJ.87.074707,PhysRevB.97.054518, PhysRevB.98.104513, PhysRevB.98.144510, PhysRevB.99.134516,vezin2019enhanced,amundsen2019quasiclassical}. Recent experimental studies show modification to the standard \textit{S}-\textit{F} proximity effect in the presence of additional spin-orbit coupling layers \cite{jeon2018enhanced, PhysRevB.97.184521, Martinez2018, PhysRevB.99.024507}. However, our own attempt to propagate an equal-spin triplet supercurrent through a Pt/Co/Ru/Co/Pt Josephson junction (with spin-orbit coupling at the Pt/Co interface) showed that, at best, the singlet-triplet conversion efficiency by spin-orbit coupling in that system is very poor \cite{satchellSOC2018}. Comparing the experimental conditions to the theoretical predictions, one major shortcoming in our previous experiment was that the in-plane \textit{F} layer chosen (a Co/Ru/Co synthetic antiferromagnet) did not satisfy the theoretical criteria that the \textit{F} layer need have magnetization both in- and out-of-plane \cite{PhysRevB.89.134517,jacobsen2016controlling}.

This work revisits our Josephson junction experiment \cite{satchellSOC2018}, employing [Co/Ni]$_n$/Co multilayers where the remanent magnetization ($M_r$) in-plane (IP) and out-of-plane (OOP) obeys the condition $M_r^{\text{IP}} \approx M_r^{\text{OOP}} \approx 0.5M_s $, where $M_s$ is the saturation magnetization. This condition can be referred to as canted magnetization. Thus, the canted [Co/Ni]$_n$/Co multilayers more closely satisfy theoretical predictions \cite{PhysRevB.89.134517,jacobsen2016controlling} than our previous work \cite{satchellSOC2018}. Josephson junctions are fabricated to compare the transport properties of; \textit{S--N--F--N--S} and \textit{S}--$N_{\text{SOC}}$--\textit{F}--$N_{\text{SOC}}$--\textit{S}, where \textit{S} is Nb, \textit{F} is the [Co/Ni]$_n$/Co multilayer, \textit{N} is Cu and $N_{\text{SOC}}$ is Pt, which has been shown in previous works to have strong Rashba spin-orbit coupling with Co due to broken inversion symmetry \cite{miron2011perpendicular,haazen2013domain,PhysRevB.90.020402}. For comparison, \textit{S--F'--F--F'--S} Josephson junctions are studied where the \textit{F'} layer Ni is known to be a good spin-mixer layer for the generation of equal-spin triplets \cite{PhysRevB.86.224506}. 

Multilayers of Co/Ni are of interest for spintronic applications such as spin-transfer torque memory (STT-MRAM) \cite{mangin2006current}, due to their favorable properties of high spin polarization (up to 90\% \cite{PhysRevMaterials.2.064410}) and large perpendicular magnetic anisotropy \cite{PhysRevB.86.014425}. In a previous Josephson junction study, this material proved to be a good candidate for propagation of equal-spin triplet supercurrent \cite{PhysRevB.86.224506}. Due to high spin-polarization in the Co/Ni system, the short-ranged supercurrent components (namely, spin-singlet and $m_{\text{s}} = 0$ triplet component) are strongly suppressed, and signatures of equal-spin triplets are clear from the greatly enhanced Josephson current.

\begin{figure*} 
\includegraphics[width=2\columnwidth]{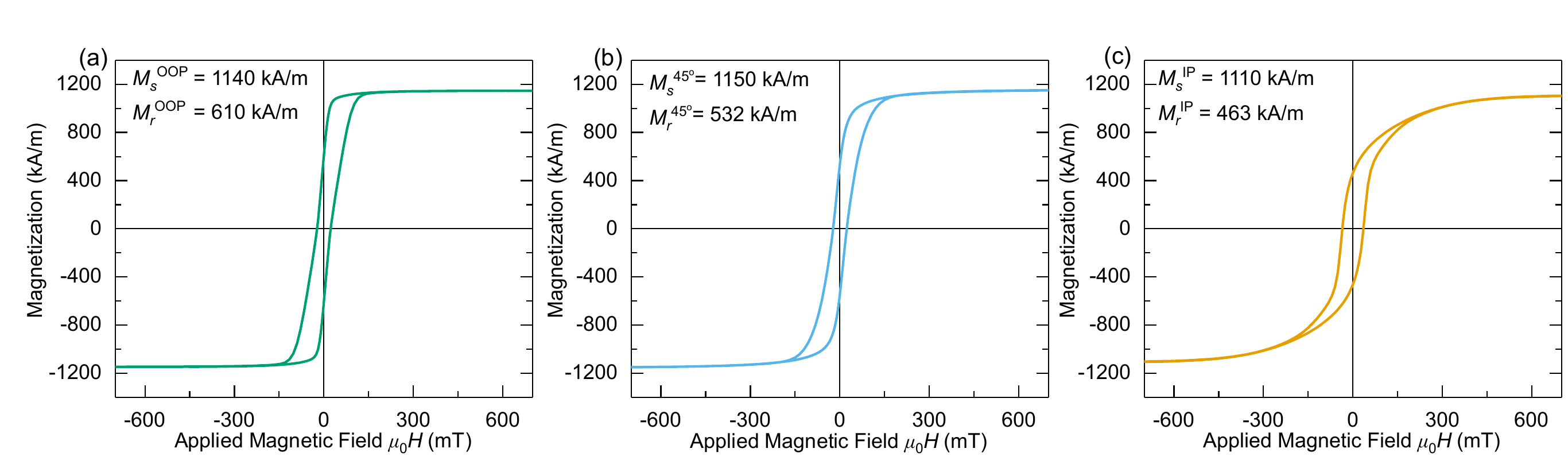} 
\caption{Magnetic characterization of the sheet film sample $S$--Pt(4.5)--[Co(0.4)/Ni(0.4)]$_{\times 8}$/Co(0.4)--Pt(4.5)--$S$. (a-c) Magnetic hysteresis loops acquired at a temperature of 10~K with the applied field oriented (a) out-of-plane, (b) at 45$^{\circ}$ to the plane and (c) in-plane. The diamagnetic contribution from the substrate has been subtracted. Values of $M$ are calculated from the measured total magnetic moments and areas of the samples, and the total nominal thicknesses of the Co and Ni layers. The uncertainty in $M$ is dominated by the area measurements (different portions of the sample were used for (a,b) and c), and is less than 5\%. (1~kA/m = 1~emu/cm$^3$). \label{loops}}
\end{figure*}

\section{Methods}

The films are deposited, patterned and measured using identical methodology to our previous work \cite{satchellSOC2018}. A Quantum Design MPMS3 magnetometer is employed to characterize sister sheet film samples at 10~K. Electrical transport is performed on patterned Josephson junctions using a conventional four-point-probe measurement configuration at 4.2~K, employing the low noise electrical transport system described in reference \cite{PhysRevB.96.224515}. Layer thicknesses (in brackets) are in nm. The bottom superconducting electrode is a multilayer [Nb(25)/Au(2.4)]$_3$/Nb(20) which grows considerably smoother than single layer Nb of comparable total thickness. The bottom electrode, layers comprising the junction, and a capping bilayer Nb(5)/Au(15) are grown without breaking vacuum. After definition of 6 and 12~$\mu$m diameter circular Josephson junctions by photolithography and ion milling, the top Au(15) layer is ion milled \textit{in situ} with the deposition of the top superconducting electrode, Nb(150).

\section{Magnetic Characterization}


Further magnetization data are available in the Supplemental Materials \cite{SM}, where careful characterization of the reorientation from out-of-plane (OOP) to in-plane (IP) magnetization is mapped in a set of samples varying the Co thickness $S$--Cu(2.5)--[Co($d_\text{Co}$)/Ni(0.4)]$_{\times 8}$/Co($d_\text{Co}$)--Cu(2.5)--$S$. It was found that $d_\text{Co}=0.2$~nm samples show strongly out-of-plane magnetization, consistent with previous work \cite{PhysRevB.86.224506}. Tuning the Co thickness allowed us to find the reorientation transition from predominant OOP to IP magnetic anisotropy. We find when $d_\text{Co}=0.4$~nm, the multilayers are at the cusp between predominant OOP and predominant IP magnetization and that by $d_\text{Co}=0.5$~nm the magnetization lies predominantly IP. For this study, we choose $d_\text{Co}=0.4$~nm for Josephson junctions which is right at the reorientation transition, since this is the best candidate for the magnetization to be canted.

The magnetization versus field data are shown in Figure~\ref{loops}, for a $S$--Pt(4.5)--[Co(0.4)/Ni(0.4)]$_{\times 8}$ /Co(0.4)--Pt(4.5)--$S$ sheet film sample at 10~K. The choice of Pt thickness here is dictated by the transport measurements to follow, although we do not find any difference in the magnetic response by varying the Pt layer thickness from 2.5 nm to 4.5 nm. For all measured applied magnetic field orientations, the sample displays hysteresis loops which are characterized as neither being typical easy axis or hard axis loops, with remanent magnetization ($M_r$) in all field orientations equal to just under half the saturation magnetization. 



The magnetization vs field data suggests the [Co(0.4)/Ni(0.4)]$_{\times 8}$/Co(0.4) multilayer cannot be described by the Stoner--Wohlfarth model of a single domain ferromagnet, since there is no obvious easy axis. Therefore the hysteresis of our samples is due to the formation of domains. Previous work suggests that the domain size of Co/Ni multilayers of similar total thickness is 100~nm \cite{MACIA20123629}. Our own characterization suggests that the size of individual domains in our samples are below the size resolution of our magnetic force microscope ($\approx 50$~nm). The measurements presented here are unable to distinguish between a remanent state where the individual domains are at a canted angle, and a remanent state where individual domains point in either the IP or OOP direction and contribute to a net magnetization in the direction of applied or set field. We believe since the $M_r$ values in each measured field orientation are so similar and the film thickness was chosen to be at the OOP-IP reorientation transition that the former case is more likely. 

\section{Electrical Transport}

\begin{figure} 
\includegraphics[width=1\columnwidth]{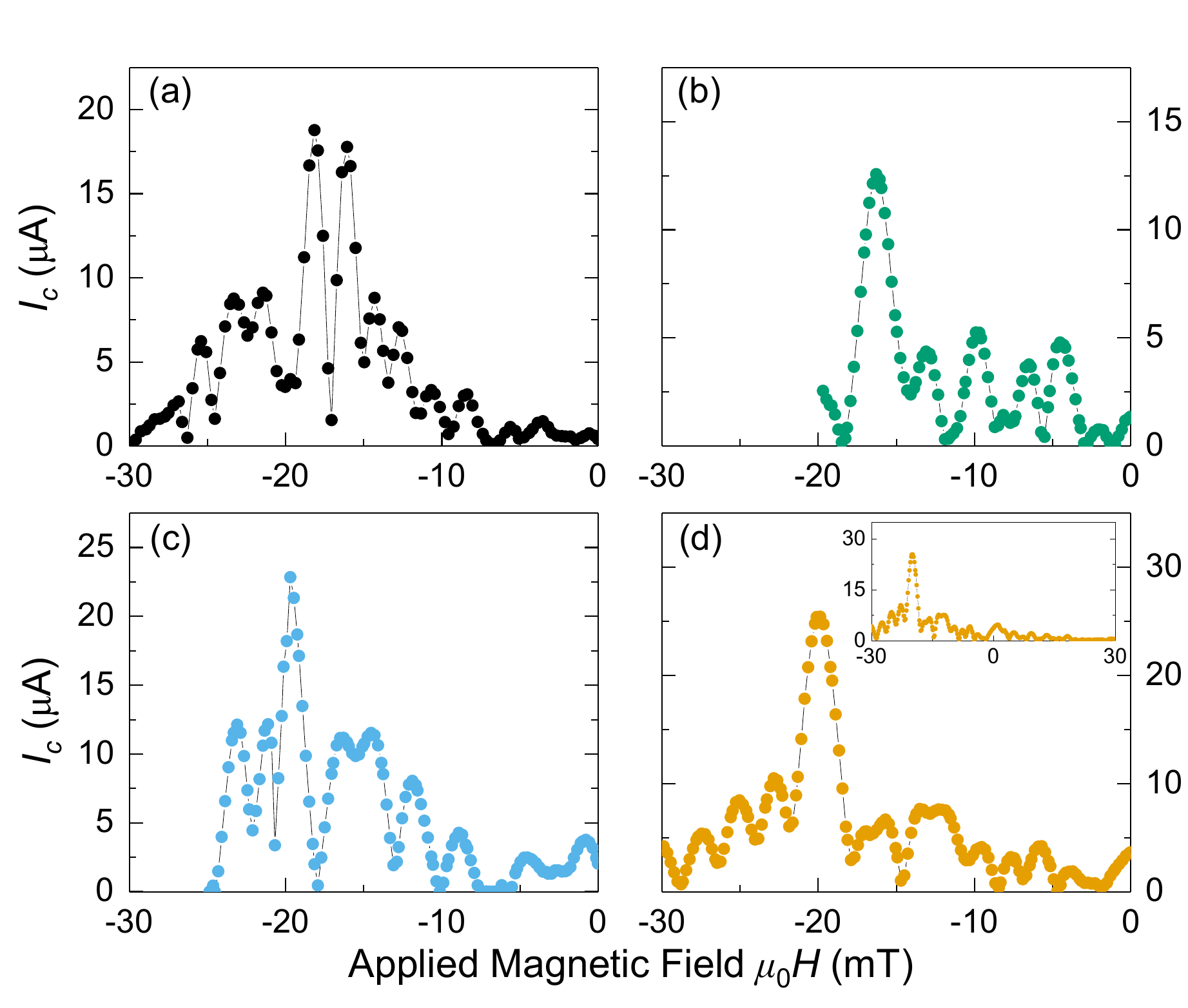} 
\caption{The critical current, $I_c$, is plotted vs the applied in-plane field for a representative circular Josephson junction of diameter 6~$\mu$m and structure \textit{S}--Pt(4.5)--[Co(0.4)/Ni(0.4)]$_{\times 8}$/Co(0.4)--Pt(4.5)--\textit{S} measured in (a) the as-grown magnetic state, (b) after \textit{ex situ} applied field of 300~mT OOP, (c) after \textit{ex situ} applied field of 300~mT at 45$^\text{o}$ to the plane, and (d) after \textit{in situ} applied field of 300~mT IP. The inset in (d) shows the same dataset over the complete measurement range. Lines through the data are a guide for the eye and the uncertainty in determining $I_c$ is smaller than the data points. \label{Juncts}}
\end{figure}


By measuring the \textit{I}--\textit{V} characteristic as function of applied magnetic field and extracting the critical current ($I_c$) assuming the resistively shunted junction model, the $I_c$($B$) `Fraunhofer' patterns for each Josephson junction are determined. Figure \ref{Juncts} shows Fraunhofer patterns for the sample \textit{S}--Pt(4.5)--[Co(0.4)/Ni(0.4)]$_{\times 8}$/Co(0.4)--Pt(4.5)--\textit{S} measured in (a) the as-grown magnetic state, (b) the magnetic state after applied OOP field of 300~mT, (c) the magnetic state after applied field of 300~mT at 45$^\text{o}$ to the plane, and (d) the magnetic state after applied IP field of 300~mT. The IP field can be applied \textit{in situ} with our measurements at 4.2~K. After the applied saturating field, we warm the sample through $T_c$ in order to remove trapped flux. The OOP and 45$^\text{o}$ fields are applied \textit{ex situ} at room temperature using a permanent magnet. Figure~\ref{loops} (d) suggests that the sample's remanent magnetization will remain aligned in the direction of the saturating field.

\begin{figure} 
\includegraphics[width=0.8\columnwidth]{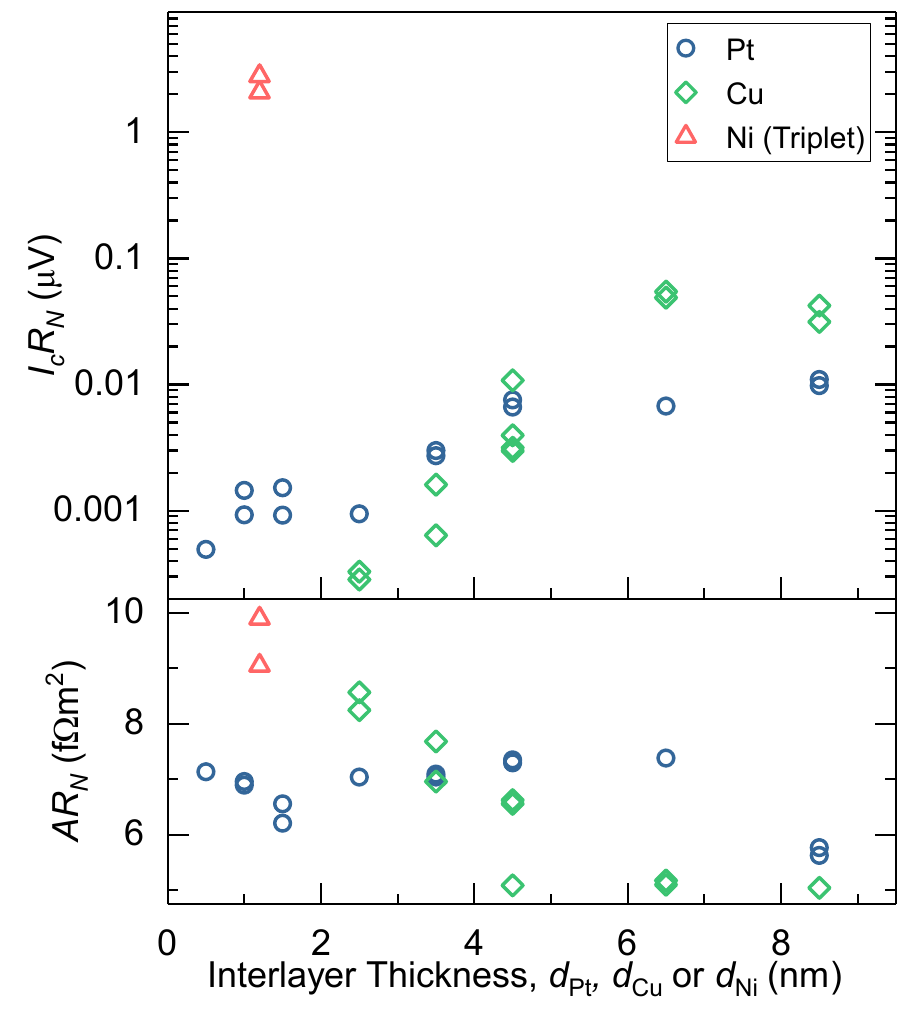} 
\caption{Top: Product of critical Josephson current times normal state resistance vs interlayer thickness ($d_{\text{Pt}}$, $d_{\text{Cu}}$ or $d_{\text{Ni}}$, (nm)) for ferromagnetic Josephson junctions of the form; $S$--$N$($d_{\text{Pt}}$ or $d_{\text{Cu}}$)--[Co(0.4)/Ni(0.4)]$_{\times 8}$/Co(0.4)--$N$($d_{\text{Pt}}$ or $d_{\text{Cu}}$)--$S$ and red triangles $S$--Cu(4.5)--Ni($d_{\text{Ni}}$)--Cu(4.5)--[Co(0.4)/Ni(0.4)]$_{\times 8}$/Co(0.4)--Cu(4.5)--Ni($d_{\text{Ni}}$)--Cu(4.5)--$S$. Bottom: Product of area times normal state resistance for the same junctions. Each data point represents one Josephson junction and the uncertainty in determining $I_cR_N$ is smaller than the data points. The scatter in $AR_N$ is most likely sample-to-sample variation in $A$.  \label{Fig3}}
\end{figure}

As Figure \ref{Juncts} shows, the as-grown state generally gives a less well defined Fraunhofer pattern due to the stray fields of the multi-domain state. Application of saturating magnetic field improves the domain structure and in some cases the peak $I_c$ value increases as a result. This $I_c^{\text{max}}$ can easily be determined for our junctions by reading from the graphs of $I_c$($B$). This value (from whichever measured magnetic state gives the largest $I_c^{\text{max}}$) is used to determine the $I_cR_N$ product of critical Josephson current times normal state resistance plotted in Figure \ref{Fig3}.

Due to the IP magnetization component of the $F$ layer, the Fraunhofer patterns are offset ($H_\text{offset}$) from $H = 0$ and the central lobe is located between about $\mu_0H_\text{offset}=-10$~mT and $\mu_0H_\text{offset}=-20$~mT, Figure \ref{Juncts}. If the magnetization of the \textit{F} layer were completely OOP, the Fraunhofer pattern would be centered about zero applied field, which is not the case here. Using the assumption that the peak of the offset Fraunhofer pattern is obtained when the applied field exactly cancels the IP component of the magnetization ($M^{\text{IP}}$), we can estimate $H_\text{offset}$ by \cite{PhysRevB.79.094523}
\begin{equation}
H_\text{offset} = \frac{-M^{\text{IP}} d_\text{F}}{2 \lambda_\text{L} + d_\text{F}},
\end{equation}
where $d_\text{F}$ is the thickness of the ferromagnet and $\lambda_\text{L}$ is the London penetration depth of the superconducting leads \cite{London}. Using the magnetization data obtained in Figure \ref{loops}; if the magnetization is completely IP ($M^{\text{IP}} = M_s^{\text{IP}}$), the Fraunhofer pattern should have $\mu_0H_\text{offset}=-47$~mT. From the transport data and the hysteresis loop, this is not the case. Clearly only a fraction of the magnetization remains in-plane due to the canting. We estimate that the offset from this fraction of the total magnetization ($M^{\text{IP}} = M_r^{\text{IP}}$) should have $\mu_0H_\text{offset}=-20$~mT, consistent with the transport data in Figure \ref{Juncts} (d). We take this as supporting evidence that the magnetization in the junctions is canted as these observations are similar to a previous Josephson junction study of the weak ferromagnet PdNi, which for large thicknesses ($>70$~nm) has both IP and OOP magnetization components \cite{PhysRevB.79.094523}.

Figure \ref{Fig3} shows the collated $I_cR_N$ and $AR_N$ (area times normal state resistance) for the Josephson junctions measured in this study. The $F$ layer for all samples is fixed as the [Co(0.4)/Ni(0.4)]$_{\times 8}$/Co(0.4) multilayer. The high consistency in $AR_N$ values between samples indicates successful and reproducible Josephson junction fabrication. Plotted are \textit{S--N--F--N--S} and \textit{S}--$N_{\text{SOC}}$--\textit{F}--$N_{\text{SOC}}$--\textit{S} where the \textit{N} interlayer Cu($d_{\text{Cu}}$) is not expected to contribute significantly to spin-orbit coupling compared to the $N_{\text{SOC}}$ Pt($d_{\text{Pt}}$) interlayer. Also shown is the result of a traditional equal-spin triplet Josephson junction, \textit{S--F'--F--F'--S} where the spin mixer \textit{F'} layers are Ni($d_{\text{Ni}}$). More completely, the triplet sample is: \textit{S--N--F'--N--F--N--F'--N--S} where the Cu(4.5) \textit{N} layers separate each active layer for the purpose of buffering the Ni growth and decoupling the magnetic switching of each ferromagnetic layer.

The results for the Pt interlayer samples can be described in three regimes. In the Pt interlayer thickness range 0.5~nm $\leq d_{\text{Pt}} \leq $ 2.5~nm the supercurrent ($I_cR_N$) has a flat dependence on $d_{\text{Pt}}$ and is small. Between 2.5~nm $\leq d_{\text{Pt}} \leq $ 4.5~nm the supercurrent is enhanced with increasing thickness. Finally, at larger thicknesses of Pt the critical current has saturated and remains constant. We would expect that upon increasing the thickness of Pt much further the $I_cR_N$ would begin to decay with $\xi_{\text{Pt}}$ (which for Pt in this work, appears to be longer than for Pt in our previous work \cite{satchellSOC2018}). The Cu interlayer samples follow a similar trend. We do not grow Cu samples thinner than 2.5~nm due to the tendency of thin Cu to grow in a nonequilibrium bcc phase on bcc Nb \cite{JACE:JACE1673,doi:10.1063/1.119611,doi:10.1063/1.371342}. The saturation in $I_cR_N$ occurs for Cu interlayers at the thicker value of Cu$=6.5$~nm, and the highest reported $I_cR_N$ for Cu is higher than any Pt sample.


\section{Discussion}\label{discussion}

The most striking feature of our experimental study is the greatly enhanced $I_cR_N$ in the \textit{S--F'--F--F'--S} Josephson junctions containing Ni(1.2) spin-mixer layers (Figure \ref{Fig3}). From the wealth of previous literature on this topic \cite{PhysRevLett.104.137002, PhysRevB.86.224506, PhysRevLett.116.077001,doi:10.1098/rsta.2015.0150}, we can say with confidence that the increased $I_cR_N$ product is due to equal-spin triplet supercurrent present in these junctions (even though we do not measure the decay length of the supercurrent in this work). The $I_cR_N$ values for these Josephson junctions are consistent with previous work on Co/Ni multilayers of comparable total thickness \cite{PhysRevB.86.224506}. 

We do not attribute any of the observations of this work to spin-orbit coupling mediating singlet-triplet conversion. Compared to the Ni(1.2) traditional \textit{S--F'--F--F'--S} triplet samples, the $I_cR_N$ product for the other Josephson junctions in this study containing Cu or Pt interlayers are between about $\times$50 and $\times$1000 lower. Even the largest Pt $I_cR_N$ is smaller than that of the Cu(6.5) sample, where we expect the contributions from spin-orbit coupling to be negligible in comparison to Pt. 

There are, however, other interesting trends in our data. The $I_cR_N$ increases dramatically with interlayer thickness for both Pt and Cu interlayer samples and then saturates within the thickness range of this study. The increase in $I_cR_N$ is larger for the Cu interlayer samples, and the peak value for Cu(6.5) represents a $\times100$ increase in $I_cR_N$ from Cu(2.5). This result is very important in the study of \textit{S}--\textit{F}--\textit{S} Josephson junctions for cryogenic memory devices. In the proposed Josephson magnetic RAM (JMRAM) memory cell architecture, the \textit{S}--\textit{F}--\textit{S} Josephson junction acts as a passive phase shifter, so must have $I_c$ greater than the \textit{S}--\textit{I}--\textit{S} junctions (where \textit{I} is an insulator) in the cell \cite{gingrich_controllable_2016,dayton2017experimental}.

The rapid increase in $I_cR_N$ with increasing Cu and Pt interlayer thickness we attribute to the type of subtle structural effects present in our previous study with these interlayers in Co/Ru/Co Josephson junctions \cite{satchellSOC2018}. In that work, we found that junctions containing fcc Pt(0.5) and fcc Cu(2.5) interlayers gave almost identical $I_cR_N$ products, which is also true in this work. However in this work, increasing Pt thickness further enhances $I_cR_N$, the opposite behavior to the previous study. The dependence on Cu thickness was not studied previously, however it is a reasonable assumption based upon other studies of Co/Ru/Co that upon increasing the Cu thickness further, $I_cR_N$ would also have increased in that system \cite{PhysRevB.80.020506}. We attributed the observations of our previous work to the fcc interlayers modifying the growth morphology of the Co layers from a mixed fcc/hcp growth without the fcc interlayers (a known phenomena creating stacking faults in the Co grains \cite{PhysRevLett.115.056601}, which may be responsible for suppressing supercurrent) to pure fcc growth with the fcc interlayers (which caused an increase in supercurrent propagation by removing the stacking faults). The increasing $I_cR_N$ with interlayer thickness in this work is most likely due to improved growth morphology of the Co/Ni multilayer by the same mechanism. We believe Co to be particularly susceptible to this phenomena; measurements in our group of junctions containing Ni$_{84}$Fe$_{16}$ do not show any change of $I_cR_N$ with Cu thickness \cite{Willard}. 

Although similar trends in $I_cR_N$ are observed here with Cu and Pt, the difference in the absolute values of $I_cR_N$ between Cu and Pt may be caused by structural differences such as; roughness, nonequilibrium growth modes \cite{JACE:JACE1673,doi:10.1063/1.119611,doi:10.1063/1.371342}, or strain caused by the different lattice parameters (Cu = 0.361 nm, Pt = 0.392 nm). Also, the magnetic moment gained by Pt in proximity to Co \cite{PhysRevB.65.020405,rowan2017interfacial}, or the different effective coherence lengths of the two proximitized normal metals could reduce $I_cR_N$ in the thickest Pt samples compared to the thickest Cu.


Finally, we compare our current and previous experiment to theoretical predictions. In our previous work \cite{satchellSOC2018} we reproduced the result of Bergeret \textit{et al.}, who consider singlet-triplet conversion in the presence of spin-orbit coupling \cite{PhysRevB.89.134517}. For completeness, we reproduce that theoretical description again. Bergeret \textit{et al.} predict that a ferromagnetic Josephson junction with Rashba and/or Dresselhaus spin-orbit coupling will propagate equal-spin triplet supercurrent under the condition that the vector operator $\big[ \mathcal{\hat{A}}_k,[ \mathcal{\hat{A}}_k, h^a \sigma^a ] \big]$ not be parallel to the exchange field operator ($h^x\sigma^x, h^y\sigma^y, h^z\sigma^z$), where the $x$ direction is OOP and $y,z$ are IP components. The complete form of $\big[ \mathcal{\hat{A}}_k,[ \mathcal{\hat{A}}_k, h^a \sigma^a ] \big]$ is shown in equation (67) of their work, which we modify here for our metallic system where the contribution to spin-orbit coupling from Rashba is non-zero ($\alpha \neq 0$) and Dresselhaus is zero ($\beta = 0$) \cite{PhysRevB.89.134517}:
\begin{equation}
\big[ \mathcal{\hat{A}}_k,[ \mathcal{\hat{A}}_k, h^a \sigma^a ] \big] = 4 \alpha^2 (2 h^x \sigma^x + h^y \sigma^y + h^z \sigma^z ),
\end{equation}
where $\sigma$ is the vector of the Pauli matrices and $\alpha$ is referred to in the literature as the Rashba constant. Equation (2) has components ($2h^x, h^y, h^z$). Therefore, in order to satisfy the condition for equal-spin triplet generation, the exchange field components $h^x$ and at least one of $h^y$ or $h^z$ must be non-zero, giving ($2h^x, h^y, h^z$) a component perpendicular to the exchange field. 

In our previous work, a major shortcoming was the predominant in-plane anisotropy ($h^x = 0$ and $h^{y \text{ or } z} \neq 0$ in the notation of the theory) of the Co/Ru/Co $F$ layers \cite{satchellSOC2018}. In the context of the theory, it is clear why that experiment should have failed to generate equal-spin triplet supercurrent as (2) has only components parallel to the exchange field. In this work we designed the study to satisfy more closely the theoretical criteria of Bergeret \textit{et al.}, since our canted ferromagnetic layers showed large remanence for IP, 45$^{\circ}$ and OOP applied fields (equivalent to $h^x \neq 0, h^{y \text{ or } z} \neq 0$) the Pt Josephson junctions in this experiment \textit{should} have produced equal-spin triplet supercurrent. A possible discrepancy between the theoretical prediction and our experiment is the multi-domain magnetic state of our $F$ layer, which may not be accounted for in the theoretical calculations. Alternatively, the efficiency of singlet-triplet conversion by spin-orbit coupling may be so poor that it falls below our experimental detection resolution, or the theoretical understanding of such systems may be incomplete.

In the future it would be interesting to try non-metallic \textit{S}--\textit{F}--\textit{S} Josephson junctions with strong intrinsic spin-orbit coupling. This could be accomplished by incorporating a semiconductor weak link with strong spin-orbit coupling such as InAs \cite{Doh272} or InSb \cite{doi:10.1021/nl303758w}. Although such a system would be difficult to fabricate in the current perpendicular to plane geometry of this work, a current in plane geometry may be possible.

\section{Conclusions} 

This work compares the role of Pt and Cu interlayers in Josephson junctions containing [Co/Ni]$_n$/Co ferromagnetic weak links, where the magnetization of the multilayer is neither in-plane nor out-of-plane. By comparing the critical current of junctions with Pt or Cu interlayers (where Pt has much larger contribution to spin-orbit coupling than Cu), we find no evidence in this work for spin-orbit coupling mediating singlet-triplet conversion. Due to both fcc Pt and fcc Cu facilitating the growth of the multilayer, a large enhancement in critical current is observed for both interlayers in the thickness range (0 - 9 nm) studied here. This result has important implications for Josephson junction devices in the field of superspintronics, where large critical currents are desirable.


The data associated with this paper are openly available from the University of Leeds data repository \cite{Dataset}.

\begin{acknowledgments}
We thank G. Burnell and F. S. Bergeret for helpful discussions, B. Bi for help with fabrication using the Keck Microfabrication Facility, and V. Aguilar for his assistance with measurement software and instrumentation. This project has received funding from the European Union’s Horizon 2020 research and innovation programme under the Marie Sk\l{}odowska-Curie grant agreement No. 743791 (SUPERSPIN). 

\end{acknowledgments}

\bibliography{library}

\begin{thebibliography}{60}%
\makeatletter
\providecommand \@ifxundefined [1]{%
 \@ifx{#1\undefined}
}%
\providecommand \@ifnum [1]{%
 \ifnum #1\expandafter \@firstoftwo
 \else \expandafter \@secondoftwo
 \fi
}%
\providecommand \@ifx [1]{%
 \ifx #1\expandafter \@firstoftwo
 \else \expandafter \@secondoftwo
 \fi
}%
\providecommand \natexlab [1]{#1}%
\providecommand \enquote  [1]{``#1''}%
\providecommand \bibnamefont  [1]{#1}%
\providecommand \bibfnamefont [1]{#1}%
\providecommand \citenamefont [1]{#1}%
\providecommand \href@noop [0]{\@secondoftwo}%
\providecommand \href [0]{\begingroup \@sanitize@url \@href}%
\providecommand \@href[1]{\@@startlink{#1}\@@href}%
\providecommand \@@href[1]{\endgroup#1\@@endlink}%
\providecommand \@sanitize@url [0]{\catcode `\\12\catcode `\$12\catcode
  `\&12\catcode `\#12\catcode `\^12\catcode `\_12\catcode `\%12\relax}%
\providecommand \@@startlink[1]{}%
\providecommand \@@endlink[0]{}%
\providecommand \url  [0]{\begingroup\@sanitize@url \@url }%
\providecommand \@url [1]{\endgroup\@href {#1}{\urlprefix }}%
\providecommand \urlprefix  [0]{URL }%
\providecommand \Eprint [0]{\href }%
\providecommand \doibase [0]{http://dx.doi.org/}%
\providecommand \selectlanguage [0]{\@gobble}%
\providecommand \bibinfo  [0]{\@secondoftwo}%
\providecommand \bibfield  [0]{\@secondoftwo}%
\providecommand \translation [1]{[#1]}%
\providecommand \BibitemOpen [0]{}%
\providecommand \bibitemStop [0]{}%
\providecommand \bibitemNoStop [0]{.\EOS\space}%
\providecommand \EOS [0]{\spacefactor3000\relax}%
\providecommand \BibitemShut  [1]{\csname bibitem#1\endcsname}%
\let\auto@bib@innerbib\@empty
\bibitem [{\citenamefont {Buzdin}(2005)}]{RevModPhys.77.935}%
  \BibitemOpen
  \bibfield  {author} {\bibinfo {author} {\bibfnamefont {A.~I.}\ \bibnamefont
  {Buzdin}},\ }\href {\doibase 10.1103/RevModPhys.77.935} {\bibfield  {journal}
  {\bibinfo  {journal} {Rev. Mod. Phys.}\ }\textbf {\bibinfo {volume} {77}},\
  \bibinfo {pages} {935} (\bibinfo {year} {2005})}\BibitemShut {NoStop}%
\bibitem [{\citenamefont {Bergeret}\ \emph {et~al.}(2005)\citenamefont
  {Bergeret}, \citenamefont {Volkov},\ and\ \citenamefont
  {Efetov}}]{RevModPhys.77.1321}%
  \BibitemOpen
  \bibfield  {author} {\bibinfo {author} {\bibfnamefont {F.~S.}\ \bibnamefont
  {Bergeret}}, \bibinfo {author} {\bibfnamefont {A.~F.}\ \bibnamefont
  {Volkov}}, \ and\ \bibinfo {author} {\bibfnamefont {K.~B.}\ \bibnamefont
  {Efetov}},\ }\href {\doibase 10.1103/RevModPhys.77.1321} {\bibfield
  {journal} {\bibinfo  {journal} {Rev. Mod. Phys.}\ }\textbf {\bibinfo {volume}
  {77}},\ \bibinfo {pages} {1321} (\bibinfo {year} {2005})}\BibitemShut
  {NoStop}%
\bibitem [{\citenamefont {Eschrig}(2011)}]{eschrig_spin-polarized_2011}%
  \BibitemOpen
  \bibfield  {author} {\bibinfo {author} {\bibfnamefont {M.}~\bibnamefont
  {Eschrig}},\ }\href {\doibase 10.1063/1.3541944} {\bibfield  {journal}
  {\bibinfo  {journal} {Phys. Today}\ }\textbf {\bibinfo {volume} {64}},\
  \bibinfo {pages} {43} (\bibinfo {year} {2011})}\BibitemShut {NoStop}%
\bibitem [{\citenamefont {Linder}\ and\ \citenamefont
  {Robinson}(2015)}]{linder_superconducting_2015}%
  \BibitemOpen
  \bibfield  {author} {\bibinfo {author} {\bibfnamefont {J.}~\bibnamefont
  {Linder}}\ and\ \bibinfo {author} {\bibfnamefont {J.~W.~A.}\ \bibnamefont
  {Robinson}},\ }\href {\doibase 10.1038/nphys3242} {\bibfield  {journal}
  {\bibinfo  {journal} {Nat. Phys.}\ }\textbf {\bibinfo {volume} {11}},\
  \bibinfo {pages} {307} (\bibinfo {year} {2015})}\BibitemShut {NoStop}%
\bibitem [{\citenamefont {Eschrig}(2015)}]{0034-4885-78-10-104501}%
  \BibitemOpen
  \bibfield  {author} {\bibinfo {author} {\bibfnamefont {M.}~\bibnamefont
  {Eschrig}},\ }\href {http://stacks.iop.org/0034-4885/78/i=10/a=104501}
  {\bibfield  {journal} {\bibinfo  {journal} {Rep. Prog. Phys}\ }\textbf
  {\bibinfo {volume} {78}},\ \bibinfo {pages} {104501} (\bibinfo {year}
  {2015})}\BibitemShut {NoStop}%
\bibitem [{\citenamefont {Birge}(2018)}]{doi:10.1098/rsta.2015.0150}%
  \BibitemOpen
  \bibfield  {author} {\bibinfo {author} {\bibfnamefont {N.~O.}\ \bibnamefont
  {Birge}},\ }\href {\doibase 10.1098/rsta.2015.0150} {\bibfield  {journal}
  {\bibinfo  {journal} {Philos. Trans. Royal Soc. A}\ }\textbf {\bibinfo
  {volume} {376}},\ \bibinfo {pages} {20150150} (\bibinfo {year}
  {2018})}\BibitemShut {NoStop}%
\bibitem [{\citenamefont {Bergeret}\ \emph {et~al.}(2001)\citenamefont
  {Bergeret}, \citenamefont {Volkov},\ and\ \citenamefont
  {Efetov}}]{PhysRevLett.86.4096}%
  \BibitemOpen
  \bibfield  {author} {\bibinfo {author} {\bibfnamefont {F.~S.}\ \bibnamefont
  {Bergeret}}, \bibinfo {author} {\bibfnamefont {A.~F.}\ \bibnamefont
  {Volkov}}, \ and\ \bibinfo {author} {\bibfnamefont {K.~B.}\ \bibnamefont
  {Efetov}},\ }\href {\doibase 10.1103/PhysRevLett.86.4096} {\bibfield
  {journal} {\bibinfo  {journal} {Phys. Rev. Lett.}\ }\textbf {\bibinfo
  {volume} {86}},\ \bibinfo {pages} {4096} (\bibinfo {year}
  {2001})}\BibitemShut {NoStop}%
\bibitem [{\citenamefont {Keizer}\ \emph {et~al.}(2006)\citenamefont {Keizer},
  \citenamefont {Goennenwein}, \citenamefont {Klapwijk}, \citenamefont {Miao},
  \citenamefont {Xiao},\ and\ \citenamefont {Gupta}}]{keizer2006spin}%
  \BibitemOpen
  \bibfield  {author} {\bibinfo {author} {\bibfnamefont {R.}~\bibnamefont
  {Keizer}}, \bibinfo {author} {\bibfnamefont {S.}~\bibnamefont {Goennenwein}},
  \bibinfo {author} {\bibfnamefont {T.}~\bibnamefont {Klapwijk}}, \bibinfo
  {author} {\bibfnamefont {G.}~\bibnamefont {Miao}}, \bibinfo {author}
  {\bibfnamefont {G.}~\bibnamefont {Xiao}}, \ and\ \bibinfo {author}
  {\bibfnamefont {A.}~\bibnamefont {Gupta}},\ }\href {\doibase
  10.1038/nature04499} {\bibfield  {journal} {\bibinfo  {journal} {Nature}\
  }\textbf {\bibinfo {volume} {439}},\ \bibinfo {pages} {825} (\bibinfo {year}
  {2006})}\BibitemShut {NoStop}%
\bibitem [{\citenamefont {Sosnin}\ \emph {et~al.}(2006)\citenamefont {Sosnin},
  \citenamefont {Cho}, \citenamefont {Petrashov},\ and\ \citenamefont
  {Volkov}}]{PhysRevLett.96.157002}%
  \BibitemOpen
  \bibfield  {author} {\bibinfo {author} {\bibfnamefont {I.}~\bibnamefont
  {Sosnin}}, \bibinfo {author} {\bibfnamefont {H.}~\bibnamefont {Cho}},
  \bibinfo {author} {\bibfnamefont {V.~T.}\ \bibnamefont {Petrashov}}, \ and\
  \bibinfo {author} {\bibfnamefont {A.~F.}\ \bibnamefont {Volkov}},\ }\href
  {\doibase 10.1103/PhysRevLett.96.157002} {\bibfield  {journal} {\bibinfo
  {journal} {Phys. Rev. Lett.}\ }\textbf {\bibinfo {volume} {96}},\ \bibinfo
  {pages} {157002} (\bibinfo {year} {2006})}\BibitemShut {NoStop}%
\bibitem [{\citenamefont {Khaire}\ \emph {et~al.}(2010)\citenamefont {Khaire},
  \citenamefont {Khasawneh}, \citenamefont {Pratt},\ and\ \citenamefont
  {Birge}}]{PhysRevLett.104.137002}%
  \BibitemOpen
  \bibfield  {author} {\bibinfo {author} {\bibfnamefont {T.~S.}\ \bibnamefont
  {Khaire}}, \bibinfo {author} {\bibfnamefont {M.~A.}\ \bibnamefont
  {Khasawneh}}, \bibinfo {author} {\bibfnamefont {W.~P.}\ \bibnamefont
  {Pratt}}, \ and\ \bibinfo {author} {\bibfnamefont {N.~O.}\ \bibnamefont
  {Birge}},\ }\href {\doibase 10.1103/PhysRevLett.104.137002} {\bibfield
  {journal} {\bibinfo  {journal} {Phys. Rev. Lett.}\ }\textbf {\bibinfo
  {volume} {104}},\ \bibinfo {pages} {137002} (\bibinfo {year}
  {2010})}\BibitemShut {NoStop}%
\bibitem [{\citenamefont {Robinson}\ \emph {et~al.}(2010)\citenamefont
  {Robinson}, \citenamefont {Witt},\ and\ \citenamefont
  {Blamire}}]{robinson2010controlled}%
  \BibitemOpen
  \bibfield  {author} {\bibinfo {author} {\bibfnamefont {J.~W.~A.}\
  \bibnamefont {Robinson}}, \bibinfo {author} {\bibfnamefont {J.~D.~S.}\
  \bibnamefont {Witt}}, \ and\ \bibinfo {author} {\bibfnamefont {M.~G.}\
  \bibnamefont {Blamire}},\ }\href {\doibase 10.1126/science.1189246}
  {\bibfield  {journal} {\bibinfo  {journal} {Science}\ }\textbf {\bibinfo
  {volume} {329}},\ \bibinfo {pages} {59} (\bibinfo {year} {2010})}\BibitemShut
  {NoStop}%
\bibitem [{\citenamefont {Anwar}\ \emph {et~al.}(2010)\citenamefont {Anwar},
  \citenamefont {Czeschka}, \citenamefont {Hesselberth}, \citenamefont
  {Porcu},\ and\ \citenamefont {Aarts}}]{PhysRevB.82.100501}%
  \BibitemOpen
  \bibfield  {author} {\bibinfo {author} {\bibfnamefont {M.~S.}\ \bibnamefont
  {Anwar}}, \bibinfo {author} {\bibfnamefont {F.}~\bibnamefont {Czeschka}},
  \bibinfo {author} {\bibfnamefont {M.}~\bibnamefont {Hesselberth}}, \bibinfo
  {author} {\bibfnamefont {M.}~\bibnamefont {Porcu}}, \ and\ \bibinfo {author}
  {\bibfnamefont {J.}~\bibnamefont {Aarts}},\ }\href {\doibase
  10.1103/PhysRevB.82.100501} {\bibfield  {journal} {\bibinfo  {journal} {Phys.
  Rev. B}\ }\textbf {\bibinfo {volume} {82}},\ \bibinfo {pages} {100501}
  (\bibinfo {year} {2010})}\BibitemShut {NoStop}%
\bibitem [{\citenamefont {Houzet}\ and\ \citenamefont
  {Buzdin}(2007)}]{PhysRevB.76.060504}%
  \BibitemOpen
  \bibfield  {author} {\bibinfo {author} {\bibfnamefont {M.}~\bibnamefont
  {Houzet}}\ and\ \bibinfo {author} {\bibfnamefont {A.~I.}\ \bibnamefont
  {Buzdin}},\ }\href {\doibase 10.1103/PhysRevB.76.060504} {\bibfield
  {journal} {\bibinfo  {journal} {Phys. Rev. B}\ }\textbf {\bibinfo {volume}
  {76}},\ \bibinfo {pages} {060504} (\bibinfo {year} {2007})}\BibitemShut
  {NoStop}%
\bibitem [{\citenamefont {Niu}(2012)}]{doi:10.1063/1.4743001}%
  \BibitemOpen
  \bibfield  {author} {\bibinfo {author} {\bibfnamefont {Z.}~\bibnamefont
  {Niu}},\ }\href {\doibase 10.1063/1.4743001} {\bibfield  {journal} {\bibinfo
  {journal} {Appl. Phys. Lett.}\ }\textbf {\bibinfo {volume} {101}},\ \bibinfo
  {pages} {062601} (\bibinfo {year} {2012})}\BibitemShut {NoStop}%
\bibitem [{\citenamefont {Bergeret}\ and\ \citenamefont
  {Tokatly}(2013)}]{PhysRevLett.110.117003}%
  \BibitemOpen
  \bibfield  {author} {\bibinfo {author} {\bibfnamefont {F.~S.}\ \bibnamefont
  {Bergeret}}\ and\ \bibinfo {author} {\bibfnamefont {I.~V.}\ \bibnamefont
  {Tokatly}},\ }\href {\doibase 10.1103/PhysRevLett.110.117003} {\bibfield
  {journal} {\bibinfo  {journal} {Phys. Rev. Lett.}\ }\textbf {\bibinfo
  {volume} {110}},\ \bibinfo {pages} {117003} (\bibinfo {year}
  {2013})}\BibitemShut {NoStop}%
\bibitem [{\citenamefont {Bergeret}\ and\ \citenamefont
  {Tokatly}(2014)}]{PhysRevB.89.134517}%
  \BibitemOpen
  \bibfield  {author} {\bibinfo {author} {\bibfnamefont {F.~S.}\ \bibnamefont
  {Bergeret}}\ and\ \bibinfo {author} {\bibfnamefont {I.~V.}\ \bibnamefont
  {Tokatly}},\ }\href {\doibase 10.1103/PhysRevB.89.134517} {\bibfield
  {journal} {\bibinfo  {journal} {Phys. Rev. B}\ }\textbf {\bibinfo {volume}
  {89}},\ \bibinfo {pages} {134517} (\bibinfo {year} {2014})}\BibitemShut
  {NoStop}%
\bibitem [{\citenamefont {Konschelle}(2014)}]{Konschelle2014}%
  \BibitemOpen
  \bibfield  {author} {\bibinfo {author} {\bibfnamefont {F.}~\bibnamefont
  {Konschelle}},\ }\href {\doibase 10.1140/epjb/e2014-50143-0} {\bibfield
  {journal} {\bibinfo  {journal} {Eur. Phys. J. B}\ }\textbf {\bibinfo {volume}
  {87}},\ \bibinfo {pages} {119} (\bibinfo {year} {2014})}\BibitemShut
  {NoStop}%
\bibitem [{\citenamefont {Jacobsen}\ \emph {et~al.}(2015)\citenamefont
  {Jacobsen}, \citenamefont {Ouassou},\ and\ \citenamefont
  {Linder}}]{PhysRevB.92.024510}%
  \BibitemOpen
  \bibfield  {author} {\bibinfo {author} {\bibfnamefont {S.~H.}\ \bibnamefont
  {Jacobsen}}, \bibinfo {author} {\bibfnamefont {J.~A.}\ \bibnamefont
  {Ouassou}}, \ and\ \bibinfo {author} {\bibfnamefont {J.}~\bibnamefont
  {Linder}},\ }\href {\doibase 10.1103/PhysRevB.92.024510} {\bibfield
  {journal} {\bibinfo  {journal} {Phys. Rev. B}\ }\textbf {\bibinfo {volume}
  {92}},\ \bibinfo {pages} {024510} (\bibinfo {year} {2015})}\BibitemShut
  {NoStop}%
\bibitem [{\citenamefont {Alidoust}\ and\ \citenamefont
  {Halterman}(2015{\natexlab{a}})}]{alidoust2015spontaneous}%
  \BibitemOpen
  \bibfield  {author} {\bibinfo {author} {\bibfnamefont {M.}~\bibnamefont
  {Alidoust}}\ and\ \bibinfo {author} {\bibfnamefont {K.}~\bibnamefont
  {Halterman}},\ }\href {\doibase 10.1088/1367-2630/17/3/033001} {\bibfield
  {journal} {\bibinfo  {journal} {New J. Phys.}\ }\textbf {\bibinfo {volume}
  {17}},\ \bibinfo {pages} {033001} (\bibinfo {year}
  {2015}{\natexlab{a}})}\BibitemShut {NoStop}%
\bibitem [{\citenamefont {Alidoust}\ and\ \citenamefont
  {Halterman}(2015{\natexlab{b}})}]{alidoust2015long}%
  \BibitemOpen
  \bibfield  {author} {\bibinfo {author} {\bibfnamefont {M.}~\bibnamefont
  {Alidoust}}\ and\ \bibinfo {author} {\bibfnamefont {K.}~\bibnamefont
  {Halterman}},\ }\href {\doibase 10.1088/0953-8984/27/23/235301} {\bibfield
  {journal} {\bibinfo  {journal} {J. Phys. Condens. Matter}\ }\textbf {\bibinfo
  {volume} {27}},\ \bibinfo {pages} {235301} (\bibinfo {year}
  {2015}{\natexlab{b}})}\BibitemShut {NoStop}%
\bibitem [{\citenamefont {Jacobsen}\ \emph {et~al.}(2016)\citenamefont
  {Jacobsen}, \citenamefont {Kulagina},\ and\ \citenamefont
  {Linder}}]{jacobsen2016controlling}%
  \BibitemOpen
  \bibfield  {author} {\bibinfo {author} {\bibfnamefont {S.~H.}\ \bibnamefont
  {Jacobsen}}, \bibinfo {author} {\bibfnamefont {I.}~\bibnamefont {Kulagina}},
  \ and\ \bibinfo {author} {\bibfnamefont {J.}~\bibnamefont {Linder}},\ }\href
  {\doibase 10.1038/srep23926} {\bibfield  {journal} {\bibinfo  {journal} {Sci.
  Rep.}\ }\textbf {\bibinfo {volume} {6}},\ \bibinfo {pages} {23926} (\bibinfo
  {year} {2016})}\BibitemShut {NoStop}%
\bibitem [{\citenamefont {Costa}\ \emph {et~al.}(2017)\citenamefont {Costa},
  \citenamefont {H\"ogl},\ and\ \citenamefont {Fabian}}]{PhysRevB.95.024514}%
  \BibitemOpen
  \bibfield  {author} {\bibinfo {author} {\bibfnamefont {A.}~\bibnamefont
  {Costa}}, \bibinfo {author} {\bibfnamefont {P.}~\bibnamefont {H\"ogl}}, \
  and\ \bibinfo {author} {\bibfnamefont {J.}~\bibnamefont {Fabian}},\ }\href
  {\doibase 10.1103/PhysRevB.95.024514} {\bibfield  {journal} {\bibinfo
  {journal} {Phys. Rev. B}\ }\textbf {\bibinfo {volume} {95}},\ \bibinfo
  {pages} {024514} (\bibinfo {year} {2017})}\BibitemShut {NoStop}%
\bibitem [{\citenamefont {Hikino}(2018)}]{doi:10.7566/JPSJ.87.074707}%
  \BibitemOpen
  \bibfield  {author} {\bibinfo {author} {\bibfnamefont {S.-I.}\ \bibnamefont
  {Hikino}},\ }\href {\doibase 10.7566/JPSJ.87.074707} {\bibfield  {journal}
  {\bibinfo  {journal} {J. Phys. Soc. Jpn}\ }\textbf {\bibinfo {volume} {87}},\
  \bibinfo {pages} {074707} (\bibinfo {year} {2018})}\BibitemShut {NoStop}%
\bibitem [{\citenamefont {Simensen}\ and\ \citenamefont
  {Linder}(2018)}]{PhysRevB.97.054518}%
  \BibitemOpen
  \bibfield  {author} {\bibinfo {author} {\bibfnamefont {H.~T.}\ \bibnamefont
  {Simensen}}\ and\ \bibinfo {author} {\bibfnamefont {J.}~\bibnamefont
  {Linder}},\ }\href {\doibase 10.1103/PhysRevB.97.054518} {\bibfield
  {journal} {\bibinfo  {journal} {Phys. Rev. B}\ }\textbf {\bibinfo {volume}
  {97}},\ \bibinfo {pages} {054518} (\bibinfo {year} {2018})}\BibitemShut
  {NoStop}%
\bibitem [{\citenamefont {Montiel}\ and\ \citenamefont
  {Eschrig}(2018)}]{PhysRevB.98.104513}%
  \BibitemOpen
  \bibfield  {author} {\bibinfo {author} {\bibfnamefont {X.}~\bibnamefont
  {Montiel}}\ and\ \bibinfo {author} {\bibfnamefont {M.}~\bibnamefont
  {Eschrig}},\ }\href {\doibase 10.1103/PhysRevB.98.104513} {\bibfield
  {journal} {\bibinfo  {journal} {Phys. Rev. B}\ }\textbf {\bibinfo {volume}
  {98}},\ \bibinfo {pages} {104513} (\bibinfo {year} {2018})}\BibitemShut
  {NoStop}%
\bibitem [{\citenamefont {Minutillo}\ \emph {et~al.}(2018)\citenamefont
  {Minutillo}, \citenamefont {Giuliano}, \citenamefont {Lucignano},
  \citenamefont {Tagliacozzo},\ and\ \citenamefont
  {Campagnano}}]{PhysRevB.98.144510}%
  \BibitemOpen
  \bibfield  {author} {\bibinfo {author} {\bibfnamefont {M.}~\bibnamefont
  {Minutillo}}, \bibinfo {author} {\bibfnamefont {D.}~\bibnamefont {Giuliano}},
  \bibinfo {author} {\bibfnamefont {P.}~\bibnamefont {Lucignano}}, \bibinfo
  {author} {\bibfnamefont {A.}~\bibnamefont {Tagliacozzo}}, \ and\ \bibinfo
  {author} {\bibfnamefont {G.}~\bibnamefont {Campagnano}},\ }\href {\doibase
  10.1103/PhysRevB.98.144510} {\bibfield  {journal} {\bibinfo  {journal} {Phys.
  Rev. B}\ }\textbf {\bibinfo {volume} {98}},\ \bibinfo {pages} {144510}
  (\bibinfo {year} {2018})}\BibitemShut {NoStop}%
\bibitem [{\citenamefont {Johnsen}\ \emph {et~al.}(2019)\citenamefont
  {Johnsen}, \citenamefont {Banerjee},\ and\ \citenamefont
  {Linder}}]{PhysRevB.99.134516}%
  \BibitemOpen
  \bibfield  {author} {\bibinfo {author} {\bibfnamefont {L.~G.}\ \bibnamefont
  {Johnsen}}, \bibinfo {author} {\bibfnamefont {N.}~\bibnamefont {Banerjee}}, \
  and\ \bibinfo {author} {\bibfnamefont {J.}~\bibnamefont {Linder}},\ }\href
  {\doibase 10.1103/PhysRevB.99.134516} {\bibfield  {journal} {\bibinfo
  {journal} {Phys. Rev. B}\ }\textbf {\bibinfo {volume} {99}},\ \bibinfo
  {pages} {134516} (\bibinfo {year} {2019})}\BibitemShut {NoStop}%
\bibitem [{\citenamefont {Vezin}\ \emph {et~al.}(2019)\citenamefont {Vezin},
  \citenamefont {Shen}, \citenamefont {Han},\ and\ \citenamefont
  {{\v{Z}}uti{\'c}}}]{vezin2019enhanced}%
  \BibitemOpen
  \bibfield  {author} {\bibinfo {author} {\bibfnamefont {T.}~\bibnamefont
  {Vezin}}, \bibinfo {author} {\bibfnamefont {C.}~\bibnamefont {Shen}},
  \bibinfo {author} {\bibfnamefont {J.~E.}\ \bibnamefont {Han}}, \ and\
  \bibinfo {author} {\bibfnamefont {I.}~\bibnamefont {{\v{Z}}uti{\'c}}},\
  }\href@noop {} {\bibfield  {journal} {\bibinfo  {journal} {arXiv preprint
  arXiv:1904.10773}\ } (\bibinfo {year} {2019})}\BibitemShut {NoStop}%
\bibitem [{\citenamefont {Amundsen}\ and\ \citenamefont
  {Linder}(2019)}]{amundsen2019quasiclassical}%
  \BibitemOpen
  \bibfield  {author} {\bibinfo {author} {\bibfnamefont {M.}~\bibnamefont
  {Amundsen}}\ and\ \bibinfo {author} {\bibfnamefont {J.}~\bibnamefont
  {Linder}},\ }\href@noop {} {\bibfield  {journal} {\bibinfo  {journal} {arXiv
  preprint arXiv:1904.11986}\ } (\bibinfo {year} {2019})}\BibitemShut {NoStop}%
\bibitem [{\citenamefont {Jeon}\ \emph {et~al.}(2018)\citenamefont {Jeon},
  \citenamefont {Ciccarelli}, \citenamefont {Ferguson}, \citenamefont
  {Kurebayashi}, \citenamefont {Cohen}, \citenamefont {Montiel}, \citenamefont
  {Eschrig}, \citenamefont {Robinson},\ and\ \citenamefont
  {Blamire}}]{jeon2018enhanced}%
  \BibitemOpen
  \bibfield  {author} {\bibinfo {author} {\bibfnamefont {K.-R.}\ \bibnamefont
  {Jeon}}, \bibinfo {author} {\bibfnamefont {C.}~\bibnamefont {Ciccarelli}},
  \bibinfo {author} {\bibfnamefont {A.~J.}\ \bibnamefont {Ferguson}}, \bibinfo
  {author} {\bibfnamefont {H.}~\bibnamefont {Kurebayashi}}, \bibinfo {author}
  {\bibfnamefont {L.~F.}\ \bibnamefont {Cohen}}, \bibinfo {author}
  {\bibfnamefont {X.}~\bibnamefont {Montiel}}, \bibinfo {author} {\bibfnamefont
  {M.}~\bibnamefont {Eschrig}}, \bibinfo {author} {\bibfnamefont {J.~W.}\
  \bibnamefont {Robinson}}, \ and\ \bibinfo {author} {\bibfnamefont {M.~G.}\
  \bibnamefont {Blamire}},\ }\href {\doibase 10.1038/s41563-018-0058-9}
  {\bibfield  {journal} {\bibinfo  {journal} {Nat. Mater.}\ }\textbf {\bibinfo
  {volume} {17}},\ \bibinfo {pages} {499} (\bibinfo {year} {2018})}\BibitemShut
  {NoStop}%
\bibitem [{\citenamefont {Banerjee}\ \emph {et~al.}(2018)\citenamefont
  {Banerjee}, \citenamefont {Ouassou}, \citenamefont {Zhu}, \citenamefont
  {Stelmashenko}, \citenamefont {Linder},\ and\ \citenamefont
  {Blamire}}]{PhysRevB.97.184521}%
  \BibitemOpen
  \bibfield  {author} {\bibinfo {author} {\bibfnamefont {N.}~\bibnamefont
  {Banerjee}}, \bibinfo {author} {\bibfnamefont {J.~A.}\ \bibnamefont
  {Ouassou}}, \bibinfo {author} {\bibfnamefont {Y.}~\bibnamefont {Zhu}},
  \bibinfo {author} {\bibfnamefont {N.~A.}\ \bibnamefont {Stelmashenko}},
  \bibinfo {author} {\bibfnamefont {J.}~\bibnamefont {Linder}}, \ and\ \bibinfo
  {author} {\bibfnamefont {M.~G.}\ \bibnamefont {Blamire}},\ }\href {\doibase
  10.1103/PhysRevB.97.184521} {\bibfield  {journal} {\bibinfo  {journal} {Phys.
  Rev. B}\ }\textbf {\bibinfo {volume} {97}},\ \bibinfo {pages} {184521}
  (\bibinfo {year} {2018})}\BibitemShut {NoStop}%
\bibitem [{\citenamefont {Mart{\'{i}}nez}\ \emph {et~al.}(2018)\citenamefont
  {Mart{\'{i}}nez}, \citenamefont {H{\"{o}}gl}, \citenamefont
  {Gonz{\'{a}}lez-Ruano}, \citenamefont {Cascales}, \citenamefont {Tiusan},
  \citenamefont {Lu}, \citenamefont {Hehn}, \citenamefont {Matos-Abiague},
  \citenamefont {Fabian}, \citenamefont {{\v{Z}}uti{\'{c}}},\ and\
  \citenamefont {Aliev}}]{Martinez2018}%
  \BibitemOpen
  \bibfield  {author} {\bibinfo {author} {\bibfnamefont {I.}~\bibnamefont
  {Mart{\'{i}}nez}}, \bibinfo {author} {\bibfnamefont {P.}~\bibnamefont
  {H{\"{o}}gl}}, \bibinfo {author} {\bibfnamefont {C.}~\bibnamefont
  {Gonz{\'{a}}lez-Ruano}}, \bibinfo {author} {\bibfnamefont {J.~P.}\
  \bibnamefont {Cascales}}, \bibinfo {author} {\bibfnamefont {C.}~\bibnamefont
  {Tiusan}}, \bibinfo {author} {\bibfnamefont {Y.}~\bibnamefont {Lu}}, \bibinfo
  {author} {\bibfnamefont {M.}~\bibnamefont {Hehn}}, \bibinfo {author}
  {\bibfnamefont {A.}~\bibnamefont {Matos-Abiague}}, \bibinfo {author}
  {\bibfnamefont {J.}~\bibnamefont {Fabian}}, \bibinfo {author} {\bibfnamefont
  {I.}~\bibnamefont {{\v{Z}}uti{\'{c}}}}, \ and\ \bibinfo {author}
  {\bibfnamefont {F.~G.}\ \bibnamefont {Aliev}},\ }\href@noop {} {\bibfield
  {journal} {\bibinfo  {journal} {arXiv preprint arXiv:1812.08090}\ } (\bibinfo
  {year} {2018})}\BibitemShut {NoStop}%
\bibitem [{\citenamefont {Jeon}\ \emph {et~al.}(2019)\citenamefont {Jeon},
  \citenamefont {Ciccarelli}, \citenamefont {Kurebayashi}, \citenamefont
  {Cohen}, \citenamefont {Montiel}, \citenamefont {Eschrig}, \citenamefont
  {Komori}, \citenamefont {Robinson},\ and\ \citenamefont
  {Blamire}}]{PhysRevB.99.024507}%
  \BibitemOpen
  \bibfield  {author} {\bibinfo {author} {\bibfnamefont {K.-R.}\ \bibnamefont
  {Jeon}}, \bibinfo {author} {\bibfnamefont {C.}~\bibnamefont {Ciccarelli}},
  \bibinfo {author} {\bibfnamefont {H.}~\bibnamefont {Kurebayashi}}, \bibinfo
  {author} {\bibfnamefont {L.~F.}\ \bibnamefont {Cohen}}, \bibinfo {author}
  {\bibfnamefont {X.}~\bibnamefont {Montiel}}, \bibinfo {author} {\bibfnamefont
  {M.}~\bibnamefont {Eschrig}}, \bibinfo {author} {\bibfnamefont
  {S.}~\bibnamefont {Komori}}, \bibinfo {author} {\bibfnamefont {J.~W.~A.}\
  \bibnamefont {Robinson}}, \ and\ \bibinfo {author} {\bibfnamefont {M.~G.}\
  \bibnamefont {Blamire}},\ }\href {\doibase 10.1103/PhysRevB.99.024507}
  {\bibfield  {journal} {\bibinfo  {journal} {Phys. Rev. B}\ }\textbf {\bibinfo
  {volume} {99}},\ \bibinfo {pages} {024507} (\bibinfo {year}
  {2019})}\BibitemShut {NoStop}%
\bibitem [{\citenamefont {Satchell}\ and\ \citenamefont
  {Birge}(2018)}]{satchellSOC2018}%
  \BibitemOpen
  \bibfield  {author} {\bibinfo {author} {\bibfnamefont {N.}~\bibnamefont
  {Satchell}}\ and\ \bibinfo {author} {\bibfnamefont {N.~O.}\ \bibnamefont
  {Birge}},\ }\href {\doibase 10.1103/PhysRevB.97.214509} {\bibfield  {journal}
  {\bibinfo  {journal} {Phys. Rev. B}\ }\textbf {\bibinfo {volume} {97}},\
  \bibinfo {pages} {214509} (\bibinfo {year} {2018})}\BibitemShut {NoStop}%
\bibitem [{\citenamefont {Miron}\ \emph {et~al.}(2011)\citenamefont {Miron},
  \citenamefont {Garello}, \citenamefont {Gaudin}, \citenamefont {Zermatten},
  \citenamefont {Costache}, \citenamefont {Auffret}, \citenamefont {Bandiera},
  \citenamefont {Rodmacq}, \citenamefont {Schuhl},\ and\ \citenamefont
  {Gambardella}}]{miron2011perpendicular}%
  \BibitemOpen
  \bibfield  {author} {\bibinfo {author} {\bibfnamefont {I.~M.}\ \bibnamefont
  {Miron}}, \bibinfo {author} {\bibfnamefont {K.}~\bibnamefont {Garello}},
  \bibinfo {author} {\bibfnamefont {G.}~\bibnamefont {Gaudin}}, \bibinfo
  {author} {\bibfnamefont {P.-J.}\ \bibnamefont {Zermatten}}, \bibinfo {author}
  {\bibfnamefont {M.~V.}\ \bibnamefont {Costache}}, \bibinfo {author}
  {\bibfnamefont {S.}~\bibnamefont {Auffret}}, \bibinfo {author} {\bibfnamefont
  {S.}~\bibnamefont {Bandiera}}, \bibinfo {author} {\bibfnamefont
  {B.}~\bibnamefont {Rodmacq}}, \bibinfo {author} {\bibfnamefont
  {A.}~\bibnamefont {Schuhl}}, \ and\ \bibinfo {author} {\bibfnamefont
  {P.}~\bibnamefont {Gambardella}},\ }\href {\doibase 10.1038/nature10309}
  {\bibfield  {journal} {\bibinfo  {journal} {Nature}\ }\textbf {\bibinfo
  {volume} {476}},\ \bibinfo {pages} {189} (\bibinfo {year}
  {2011})}\BibitemShut {NoStop}%
\bibitem [{\citenamefont {Haazen}\ \emph {et~al.}(2013)\citenamefont {Haazen},
  \citenamefont {Mur{\`e}}, \citenamefont {Franken}, \citenamefont {Lavrijsen},
  \citenamefont {Swagten},\ and\ \citenamefont {Koopmans}}]{haazen2013domain}%
  \BibitemOpen
  \bibfield  {author} {\bibinfo {author} {\bibfnamefont {P.}~\bibnamefont
  {Haazen}}, \bibinfo {author} {\bibfnamefont {E.}~\bibnamefont {Mur{\`e}}},
  \bibinfo {author} {\bibfnamefont {J.}~\bibnamefont {Franken}}, \bibinfo
  {author} {\bibfnamefont {R.}~\bibnamefont {Lavrijsen}}, \bibinfo {author}
  {\bibfnamefont {H.}~\bibnamefont {Swagten}}, \ and\ \bibinfo {author}
  {\bibfnamefont {B.}~\bibnamefont {Koopmans}},\ }\href {\doibase
  10.1038/nmat3553} {\bibfield  {journal} {\bibinfo  {journal} {Nat. Mater.}\
  }\textbf {\bibinfo {volume} {12}},\ \bibinfo {pages} {299} (\bibinfo {year}
  {2013})}\BibitemShut {NoStop}%
\bibitem [{\citenamefont {Hrabec}\ \emph {et~al.}(2014)\citenamefont {Hrabec},
  \citenamefont {Porter}, \citenamefont {Wells}, \citenamefont {Benitez},
  \citenamefont {Burnell}, \citenamefont {McVitie}, \citenamefont {McGrouther},
  \citenamefont {Moore},\ and\ \citenamefont {Marrows}}]{PhysRevB.90.020402}%
  \BibitemOpen
  \bibfield  {author} {\bibinfo {author} {\bibfnamefont {A.}~\bibnamefont
  {Hrabec}}, \bibinfo {author} {\bibfnamefont {N.~A.}\ \bibnamefont {Porter}},
  \bibinfo {author} {\bibfnamefont {A.}~\bibnamefont {Wells}}, \bibinfo
  {author} {\bibfnamefont {M.~J.}\ \bibnamefont {Benitez}}, \bibinfo {author}
  {\bibfnamefont {G.}~\bibnamefont {Burnell}}, \bibinfo {author} {\bibfnamefont
  {S.}~\bibnamefont {McVitie}}, \bibinfo {author} {\bibfnamefont
  {D.}~\bibnamefont {McGrouther}}, \bibinfo {author} {\bibfnamefont {T.~A.}\
  \bibnamefont {Moore}}, \ and\ \bibinfo {author} {\bibfnamefont {C.~H.}\
  \bibnamefont {Marrows}},\ }\href {\doibase 10.1103/PhysRevB.90.020402}
  {\bibfield  {journal} {\bibinfo  {journal} {Phys. Rev. B}\ }\textbf {\bibinfo
  {volume} {90}},\ \bibinfo {pages} {020402} (\bibinfo {year}
  {2014})}\BibitemShut {NoStop}%
\bibitem [{\citenamefont {Gingrich}\ \emph {et~al.}(2012)\citenamefont
  {Gingrich}, \citenamefont {Quarterman}, \citenamefont {Wang}, \citenamefont
  {Loloee}, \citenamefont {Pratt},\ and\ \citenamefont
  {Birge}}]{PhysRevB.86.224506}%
  \BibitemOpen
  \bibfield  {author} {\bibinfo {author} {\bibfnamefont {E.~C.}\ \bibnamefont
  {Gingrich}}, \bibinfo {author} {\bibfnamefont {P.}~\bibnamefont
  {Quarterman}}, \bibinfo {author} {\bibfnamefont {Y.}~\bibnamefont {Wang}},
  \bibinfo {author} {\bibfnamefont {R.}~\bibnamefont {Loloee}}, \bibinfo
  {author} {\bibfnamefont {W.~P.}\ \bibnamefont {Pratt}}, \ and\ \bibinfo
  {author} {\bibfnamefont {N.~O.}\ \bibnamefont {Birge}},\ }\href {\doibase
  10.1103/PhysRevB.86.224506} {\bibfield  {journal} {\bibinfo  {journal} {Phys.
  Rev. B}\ }\textbf {\bibinfo {volume} {86}},\ \bibinfo {pages} {224506}
  (\bibinfo {year} {2012})}\BibitemShut {NoStop}%
\bibitem [{\citenamefont {Mangin}\ \emph {et~al.}(2006)\citenamefont {Mangin},
  \citenamefont {Ravelosona}, \citenamefont {Katine}, \citenamefont {Carey},
  \citenamefont {Terris},\ and\ \citenamefont {Fullerton}}]{mangin2006current}%
  \BibitemOpen
  \bibfield  {author} {\bibinfo {author} {\bibfnamefont {S.}~\bibnamefont
  {Mangin}}, \bibinfo {author} {\bibfnamefont {D.}~\bibnamefont {Ravelosona}},
  \bibinfo {author} {\bibfnamefont {J.}~\bibnamefont {Katine}}, \bibinfo
  {author} {\bibfnamefont {M.}~\bibnamefont {Carey}}, \bibinfo {author}
  {\bibfnamefont {B.}~\bibnamefont {Terris}}, \ and\ \bibinfo {author}
  {\bibfnamefont {E.~E.}\ \bibnamefont {Fullerton}},\ }\href {\doibase
  10.1038/nmat1595} {\bibfield  {journal} {\bibinfo  {journal} {Nature
  materials}\ }\textbf {\bibinfo {volume} {5}},\ \bibinfo {pages} {210}
  (\bibinfo {year} {2006})}\BibitemShut {NoStop}%
\bibitem [{\citenamefont {Andrieu}\ \emph {et~al.}(2018)\citenamefont
  {Andrieu}, \citenamefont {Hauet}, \citenamefont {Gottwald}, \citenamefont
  {Rajanikanth}, \citenamefont {Calmels}, \citenamefont {Bataille},
  \citenamefont {Montaigne}, \citenamefont {Mangin}, \citenamefont {Otero},
  \citenamefont {Ohresser}, \citenamefont {Le~F\`evre}, \citenamefont
  {Bertran}, \citenamefont {Resta}, \citenamefont {Vlad}, \citenamefont
  {Coati},\ and\ \citenamefont {Garreau}}]{PhysRevMaterials.2.064410}%
  \BibitemOpen
  \bibfield  {author} {\bibinfo {author} {\bibfnamefont {S.}~\bibnamefont
  {Andrieu}}, \bibinfo {author} {\bibfnamefont {T.}~\bibnamefont {Hauet}},
  \bibinfo {author} {\bibfnamefont {M.}~\bibnamefont {Gottwald}}, \bibinfo
  {author} {\bibfnamefont {A.}~\bibnamefont {Rajanikanth}}, \bibinfo {author}
  {\bibfnamefont {L.}~\bibnamefont {Calmels}}, \bibinfo {author} {\bibfnamefont
  {A.~M.}\ \bibnamefont {Bataille}}, \bibinfo {author} {\bibfnamefont
  {F.}~\bibnamefont {Montaigne}}, \bibinfo {author} {\bibfnamefont
  {S.}~\bibnamefont {Mangin}}, \bibinfo {author} {\bibfnamefont
  {E.}~\bibnamefont {Otero}}, \bibinfo {author} {\bibfnamefont
  {P.}~\bibnamefont {Ohresser}}, \bibinfo {author} {\bibfnamefont
  {P.}~\bibnamefont {Le~F\`evre}}, \bibinfo {author} {\bibfnamefont
  {F.}~\bibnamefont {Bertran}}, \bibinfo {author} {\bibfnamefont
  {A.}~\bibnamefont {Resta}}, \bibinfo {author} {\bibfnamefont
  {A.}~\bibnamefont {Vlad}}, \bibinfo {author} {\bibfnamefont {A.}~\bibnamefont
  {Coati}}, \ and\ \bibinfo {author} {\bibfnamefont {Y.}~\bibnamefont
  {Garreau}},\ }\href {\doibase 10.1103/PhysRevMaterials.2.064410} {\bibfield
  {journal} {\bibinfo  {journal} {Phys. Rev. Materials}\ }\textbf {\bibinfo
  {volume} {2}},\ \bibinfo {pages} {064410} (\bibinfo {year}
  {2018})}\BibitemShut {NoStop}%
\bibitem [{\citenamefont {Gottwald}\ \emph {et~al.}(2012)\citenamefont
  {Gottwald}, \citenamefont {Andrieu}, \citenamefont {Gimbert}, \citenamefont
  {Shipton}, \citenamefont {Calmels}, \citenamefont {Magen}, \citenamefont
  {Snoeck}, \citenamefont {Liberati}, \citenamefont {Hauet}, \citenamefont
  {Arenholz}, \citenamefont {Mangin},\ and\ \citenamefont
  {Fullerton}}]{PhysRevB.86.014425}%
  \BibitemOpen
  \bibfield  {author} {\bibinfo {author} {\bibfnamefont {M.}~\bibnamefont
  {Gottwald}}, \bibinfo {author} {\bibfnamefont {S.}~\bibnamefont {Andrieu}},
  \bibinfo {author} {\bibfnamefont {F.}~\bibnamefont {Gimbert}}, \bibinfo
  {author} {\bibfnamefont {E.}~\bibnamefont {Shipton}}, \bibinfo {author}
  {\bibfnamefont {L.}~\bibnamefont {Calmels}}, \bibinfo {author} {\bibfnamefont
  {C.}~\bibnamefont {Magen}}, \bibinfo {author} {\bibfnamefont
  {E.}~\bibnamefont {Snoeck}}, \bibinfo {author} {\bibfnamefont
  {M.}~\bibnamefont {Liberati}}, \bibinfo {author} {\bibfnamefont
  {T.}~\bibnamefont {Hauet}}, \bibinfo {author} {\bibfnamefont
  {E.}~\bibnamefont {Arenholz}}, \bibinfo {author} {\bibfnamefont
  {S.}~\bibnamefont {Mangin}}, \ and\ \bibinfo {author} {\bibfnamefont {E.~E.}\
  \bibnamefont {Fullerton}},\ }\href {\doibase 10.1103/PhysRevB.86.014425}
  {\bibfield  {journal} {\bibinfo  {journal} {Phys. Rev. B}\ }\textbf {\bibinfo
  {volume} {86}},\ \bibinfo {pages} {014425} (\bibinfo {year}
  {2012})}\BibitemShut {NoStop}%
\bibitem [{\citenamefont {Glick}\ \emph {et~al.}(2017)\citenamefont {Glick},
  \citenamefont {Edwards}, \citenamefont {Korucu}, \citenamefont {Aguilar},
  \citenamefont {Niedzielski}, \citenamefont {Loloee}, \citenamefont {Pratt},
  \citenamefont {Birge}, \citenamefont {Kotula},\ and\ \citenamefont
  {Missert}}]{PhysRevB.96.224515}%
  \BibitemOpen
  \bibfield  {author} {\bibinfo {author} {\bibfnamefont {J.~A.}\ \bibnamefont
  {Glick}}, \bibinfo {author} {\bibfnamefont {S.}~\bibnamefont {Edwards}},
  \bibinfo {author} {\bibfnamefont {D.}~\bibnamefont {Korucu}}, \bibinfo
  {author} {\bibfnamefont {V.}~\bibnamefont {Aguilar}}, \bibinfo {author}
  {\bibfnamefont {B.~M.}\ \bibnamefont {Niedzielski}}, \bibinfo {author}
  {\bibfnamefont {R.}~\bibnamefont {Loloee}}, \bibinfo {author} {\bibfnamefont
  {W.~P.}\ \bibnamefont {Pratt}}, \bibinfo {author} {\bibfnamefont {N.~O.}\
  \bibnamefont {Birge}}, \bibinfo {author} {\bibfnamefont {P.~G.}\ \bibnamefont
  {Kotula}}, \ and\ \bibinfo {author} {\bibfnamefont {N.}~\bibnamefont
  {Missert}},\ }\href {\doibase 10.1103/PhysRevB.96.224515} {\bibfield
  {journal} {\bibinfo  {journal} {Phys. Rev. B}\ }\textbf {\bibinfo {volume}
  {96}},\ \bibinfo {pages} {224515} (\bibinfo {year} {2017})}\BibitemShut
  {NoStop}%
\bibitem [{SM()}]{SM}%
  \BibitemOpen
  \href@noop {} {}\bibinfo {note} {{See Supplemental Material at [URL will be
  inserted by publisher] for additional magnetic
  characterization.}}\BibitemShut {Stop}%
\bibitem [{\citenamefont {Maci{\`a}}\ \emph {et~al.}(2012)\citenamefont
  {Maci{\`a}}, \citenamefont {Warnicke}, \citenamefont {Bedau}, \citenamefont
  {Im}, \citenamefont {Fischer}, \citenamefont {Arena},\ and\ \citenamefont
  {Kent}}]{MACIA20123629}%
  \BibitemOpen
  \bibfield  {author} {\bibinfo {author} {\bibfnamefont {F.}~\bibnamefont
  {Maci{\`a}}}, \bibinfo {author} {\bibfnamefont {P.}~\bibnamefont {Warnicke}},
  \bibinfo {author} {\bibfnamefont {D.}~\bibnamefont {Bedau}}, \bibinfo
  {author} {\bibfnamefont {M.-Y.}\ \bibnamefont {Im}}, \bibinfo {author}
  {\bibfnamefont {P.}~\bibnamefont {Fischer}}, \bibinfo {author} {\bibfnamefont
  {D.}~\bibnamefont {Arena}}, \ and\ \bibinfo {author} {\bibfnamefont
  {A.}~\bibnamefont {Kent}},\ }\href {\doibase
  https://doi.org/10.1016/j.jmmm.2012.03.063} {\bibfield  {journal} {\bibinfo
  {journal} {J. Magn. Magn. Mater}\ }\textbf {\bibinfo {volume} {324}},\
  \bibinfo {pages} {3629 } (\bibinfo {year} {2012})}\BibitemShut {NoStop}%
\bibitem [{\citenamefont {Khaire}\ \emph {et~al.}(2009)\citenamefont {Khaire},
  \citenamefont {Pratt},\ and\ \citenamefont {Birge}}]{PhysRevB.79.094523}%
  \BibitemOpen
  \bibfield  {author} {\bibinfo {author} {\bibfnamefont {T.~S.}\ \bibnamefont
  {Khaire}}, \bibinfo {author} {\bibfnamefont {W.~P.}\ \bibnamefont {Pratt}}, \
  and\ \bibinfo {author} {\bibfnamefont {N.~O.}\ \bibnamefont {Birge}},\ }\href
  {\doibase 10.1103/PhysRevB.79.094523} {\bibfield  {journal} {\bibinfo
  {journal} {Phys. Rev. B}\ }\textbf {\bibinfo {volume} {79}},\ \bibinfo
  {pages} {094523} (\bibinfo {year} {2009})}\BibitemShut {NoStop}%
\bibitem [{Lon()}]{London}%
  \BibitemOpen
  \href@noop {} {}\bibinfo {note} {{Typically the bottom and top
  superconducting electrodes have identical London penetration depths, however
  in our case the bottom electrode is a Nb/Au multilayer with $\lambda_\text{L}
  =110$~nm and the top electrode is single layer Nb with
  $\lambda_\text{L}=85$~nm. When the value of $\lambda_\text{L}$ is comparable
  to the thickness of the superconductor ($d_S$), $\lambda_\text{L}$ in
  equation (1) should be replaced by $\lambda_\text{L}
  \tanh{(d_S/2\lambda_\text{L})}$, see A. Barone and G. Patern\`{o}
  \textit{Physics and Applications of the Josephson Effect} (John Wiley \&
  Sons, 1982). These considerations are taken into account in our calculations
  }}\BibitemShut {NoStop}%
\bibitem [{\citenamefont {Mitchell}\ \emph {et~al.}(1997)\citenamefont
  {Mitchell}, \citenamefont {Lu}, \citenamefont {Griffin}, \citenamefont
  {Nastasi},\ and\ \citenamefont {Kung}}]{JACE:JACE1673}%
  \BibitemOpen
  \bibfield  {author} {\bibinfo {author} {\bibfnamefont {T.~E.}\ \bibnamefont
  {Mitchell}}, \bibinfo {author} {\bibfnamefont {Y.~C.}\ \bibnamefont {Lu}},
  \bibinfo {author} {\bibfnamefont {A.~J.}\ \bibnamefont {Griffin}}, \bibinfo
  {author} {\bibfnamefont {M.}~\bibnamefont {Nastasi}}, \ and\ \bibinfo
  {author} {\bibfnamefont {H.}~\bibnamefont {Kung}},\ }\href {\doibase
  10.1111/j.1151-2916.1997.tb03037.x} {\bibfield  {journal} {\bibinfo
  {journal} {J. Am. Ceram. Soc.}\ }\textbf {\bibinfo {volume} {80}},\ \bibinfo
  {pages} {1673} (\bibinfo {year} {1997})}\BibitemShut {NoStop}%
\bibitem [{\citenamefont {Kung}\ \emph {et~al.}(1997)\citenamefont {Kung},
  \citenamefont {Lu}, \citenamefont {Griffin.}, \citenamefont {Nastasi},
  \citenamefont {Mitchell},\ and\ \citenamefont
  {Embury}}]{doi:10.1063/1.119611}%
  \BibitemOpen
  \bibfield  {author} {\bibinfo {author} {\bibfnamefont {H.}~\bibnamefont
  {Kung}}, \bibinfo {author} {\bibfnamefont {Y.-C.}\ \bibnamefont {Lu}},
  \bibinfo {author} {\bibfnamefont {A.~J.}\ \bibnamefont {Griffin.}}, \bibinfo
  {author} {\bibfnamefont {M.}~\bibnamefont {Nastasi}}, \bibinfo {author}
  {\bibfnamefont {T.~E.}\ \bibnamefont {Mitchell}}, \ and\ \bibinfo {author}
  {\bibfnamefont {J.~D.}\ \bibnamefont {Embury}},\ }\href {\doibase
  10.1063/1.119611} {\bibfield  {journal} {\bibinfo  {journal} {Appl. Phys.
  Lett.}\ }\textbf {\bibinfo {volume} {71}},\ \bibinfo {pages} {2103} (\bibinfo
  {year} {1997})}\BibitemShut {NoStop}%
\bibitem [{\citenamefont {Geng}\ \emph {et~al.}(1999)\citenamefont {Geng},
  \citenamefont {Heckman}, \citenamefont {Pratt}, \citenamefont {Bass},
  \citenamefont {Espinosa}, \citenamefont {Conradson}, \citenamefont
  {Lederman},\ and\ \citenamefont {Crimp}}]{doi:10.1063/1.371342}%
  \BibitemOpen
  \bibfield  {author} {\bibinfo {author} {\bibfnamefont {H.}~\bibnamefont
  {Geng}}, \bibinfo {author} {\bibfnamefont {J.~W.}\ \bibnamefont {Heckman}},
  \bibinfo {author} {\bibfnamefont {W.~P.}\ \bibnamefont {Pratt}}, \bibinfo
  {author} {\bibfnamefont {J.}~\bibnamefont {Bass}}, \bibinfo {author}
  {\bibfnamefont {F.~J.}\ \bibnamefont {Espinosa}}, \bibinfo {author}
  {\bibfnamefont {S.~D.}\ \bibnamefont {Conradson}}, \bibinfo {author}
  {\bibfnamefont {D.}~\bibnamefont {Lederman}}, \ and\ \bibinfo {author}
  {\bibfnamefont {M.~A.}\ \bibnamefont {Crimp}},\ }\href {\doibase
  10.1063/1.371342} {\bibfield  {journal} {\bibinfo  {journal} {J. Appl.
  Phys.}\ }\textbf {\bibinfo {volume} {86}},\ \bibinfo {pages} {4166} (\bibinfo
  {year} {1999})}\BibitemShut {NoStop}%
\bibitem [{\citenamefont {Martinez}\ \emph {et~al.}(2016)\citenamefont
  {Martinez}, \citenamefont {Pratt},\ and\ \citenamefont
  {Birge}}]{PhysRevLett.116.077001}%
  \BibitemOpen
  \bibfield  {author} {\bibinfo {author} {\bibfnamefont {W.~M.}\ \bibnamefont
  {Martinez}}, \bibinfo {author} {\bibfnamefont {W.~P.}\ \bibnamefont {Pratt}},
  \ and\ \bibinfo {author} {\bibfnamefont {N.~O.}\ \bibnamefont {Birge}},\
  }\href {\doibase 10.1103/PhysRevLett.116.077001} {\bibfield  {journal}
  {\bibinfo  {journal} {Phys. Rev. Lett.}\ }\textbf {\bibinfo {volume} {116}},\
  \bibinfo {pages} {077001} (\bibinfo {year} {2016})}\BibitemShut {NoStop}%
\bibitem [{\citenamefont {Gingrich}\ \emph {et~al.}(2016)\citenamefont
  {Gingrich}, \citenamefont {Niedzielski}, \citenamefont {Glick}, \citenamefont
  {Wang}, \citenamefont {Miller}, \citenamefont {Loloee}, \citenamefont
  {Pratt},\ and\ \citenamefont {Birge}}]{gingrich_controllable_2016}%
  \BibitemOpen
  \bibfield  {author} {\bibinfo {author} {\bibfnamefont {E.~C.}\ \bibnamefont
  {Gingrich}}, \bibinfo {author} {\bibfnamefont {B.~M.}\ \bibnamefont
  {Niedzielski}}, \bibinfo {author} {\bibfnamefont {J.~A.}\ \bibnamefont
  {Glick}}, \bibinfo {author} {\bibfnamefont {Y.}~\bibnamefont {Wang}},
  \bibinfo {author} {\bibfnamefont {D.~L.}\ \bibnamefont {Miller}}, \bibinfo
  {author} {\bibfnamefont {R.}~\bibnamefont {Loloee}}, \bibinfo {author}
  {\bibfnamefont {W.~P.}\ \bibnamefont {Pratt}}, \ and\ \bibinfo {author}
  {\bibfnamefont {N.~O.}\ \bibnamefont {Birge}},\ }\href {\doibase
  doi:10.1038/nphys3681} {\bibfield  {journal} {\bibinfo  {journal} {Nat.
  Phys.}\ }\textbf {\bibinfo {volume} {12}},\ \bibinfo {pages} {564} (\bibinfo
  {year} {2016})}\BibitemShut {NoStop}%
\bibitem [{\citenamefont {Dayton}\ \emph {et~al.}(2018)\citenamefont {Dayton},
  \citenamefont {Sage}, \citenamefont {Gingrich}, \citenamefont {Loving},
  \citenamefont {Ambrose}, \citenamefont {Siwak}, \citenamefont {Keebaugh},
  \citenamefont {Kirby}, \citenamefont {Miller}, \citenamefont {Herr},
  \citenamefont {Herr},\ and\ \citenamefont {Naaman}}]{dayton2017experimental}%
  \BibitemOpen
  \bibfield  {author} {\bibinfo {author} {\bibfnamefont {I.}~\bibnamefont
  {Dayton}}, \bibinfo {author} {\bibfnamefont {T.}~\bibnamefont {Sage}},
  \bibinfo {author} {\bibfnamefont {E.}~\bibnamefont {Gingrich}}, \bibinfo
  {author} {\bibfnamefont {M.}~\bibnamefont {Loving}}, \bibinfo {author}
  {\bibfnamefont {T.}~\bibnamefont {Ambrose}}, \bibinfo {author} {\bibfnamefont
  {N.}~\bibnamefont {Siwak}}, \bibinfo {author} {\bibfnamefont
  {S.}~\bibnamefont {Keebaugh}}, \bibinfo {author} {\bibfnamefont
  {C.}~\bibnamefont {Kirby}}, \bibinfo {author} {\bibfnamefont
  {D.}~\bibnamefont {Miller}}, \bibinfo {author} {\bibfnamefont
  {A.}~\bibnamefont {Herr}}, \bibinfo {author} {\bibfnamefont {Q.}~\bibnamefont
  {Herr}}, \ and\ \bibinfo {author} {\bibfnamefont {O.}~\bibnamefont
  {Naaman}},\ }\href {\doibase 10.1109/LMAG.2018.2801820} {\bibfield  {journal}
  {\bibinfo  {journal} {IEEE Magn. Lett.}\ }\textbf {\bibinfo {volume} {9}},\
  \bibinfo {pages} {3301905} (\bibinfo {year} {2018})}\BibitemShut {NoStop}%
\bibitem [{\citenamefont {Khasawneh}\ \emph {et~al.}(2009)\citenamefont
  {Khasawneh}, \citenamefont {Pratt},\ and\ \citenamefont
  {Birge}}]{PhysRevB.80.020506}%
  \BibitemOpen
  \bibfield  {author} {\bibinfo {author} {\bibfnamefont {M.~A.}\ \bibnamefont
  {Khasawneh}}, \bibinfo {author} {\bibfnamefont {W.~P.}\ \bibnamefont
  {Pratt}}, \ and\ \bibinfo {author} {\bibfnamefont {N.~O.}\ \bibnamefont
  {Birge}},\ }\href {\doibase 10.1103/PhysRevB.80.020506} {\bibfield  {journal}
  {\bibinfo  {journal} {Phys. Rev. B}\ }\textbf {\bibinfo {volume} {80}},\
  \bibinfo {pages} {020506} (\bibinfo {year} {2009})}\BibitemShut {NoStop}%
\bibitem [{\citenamefont {Toka\ifmmode~\mbox{\c{c}}\else \c{c}\fi{}}\ \emph
  {et~al.}(2015)\citenamefont {Toka\ifmmode~\mbox{\c{c}}\else \c{c}\fi{}},
  \citenamefont {Bunyaev}, \citenamefont {Kakazei}, \citenamefont {Schmool},
  \citenamefont {Atkinson},\ and\ \citenamefont
  {Hindmarch}}]{PhysRevLett.115.056601}%
  \BibitemOpen
  \bibfield  {author} {\bibinfo {author} {\bibfnamefont {M.}~\bibnamefont
  {Toka\ifmmode~\mbox{\c{c}}\else \c{c}\fi{}}}, \bibinfo {author}
  {\bibfnamefont {S.~A.}\ \bibnamefont {Bunyaev}}, \bibinfo {author}
  {\bibfnamefont {G.~N.}\ \bibnamefont {Kakazei}}, \bibinfo {author}
  {\bibfnamefont {D.~S.}\ \bibnamefont {Schmool}}, \bibinfo {author}
  {\bibfnamefont {D.}~\bibnamefont {Atkinson}}, \ and\ \bibinfo {author}
  {\bibfnamefont {A.~T.}\ \bibnamefont {Hindmarch}},\ }\href {\doibase
  10.1103/PhysRevLett.115.056601} {\bibfield  {journal} {\bibinfo  {journal}
  {Phys. Rev. Lett.}\ }\textbf {\bibinfo {volume} {115}},\ \bibinfo {pages}
  {056601} (\bibinfo {year} {2015})}\BibitemShut {NoStop}%
\bibitem [{Wil()}]{Willard}%
  \BibitemOpen
  \href@noop {} {}\bibinfo {note} {{J. C. Willard \textit{et al.} In
  preparation}}\BibitemShut {NoStop}%
\bibitem [{\citenamefont {Geissler}\ \emph {et~al.}(2001)\citenamefont
  {Geissler}, \citenamefont {Goering}, \citenamefont {Justen}, \citenamefont
  {Weigand}, \citenamefont {Sch\"utz}, \citenamefont {Langer}, \citenamefont
  {Schmitz}, \citenamefont {Maletta},\ and\ \citenamefont
  {Mattheis}}]{PhysRevB.65.020405}%
  \BibitemOpen
  \bibfield  {author} {\bibinfo {author} {\bibfnamefont {J.}~\bibnamefont
  {Geissler}}, \bibinfo {author} {\bibfnamefont {E.}~\bibnamefont {Goering}},
  \bibinfo {author} {\bibfnamefont {M.}~\bibnamefont {Justen}}, \bibinfo
  {author} {\bibfnamefont {F.}~\bibnamefont {Weigand}}, \bibinfo {author}
  {\bibfnamefont {G.}~\bibnamefont {Sch\"utz}}, \bibinfo {author}
  {\bibfnamefont {J.}~\bibnamefont {Langer}}, \bibinfo {author} {\bibfnamefont
  {D.}~\bibnamefont {Schmitz}}, \bibinfo {author} {\bibfnamefont
  {H.}~\bibnamefont {Maletta}}, \ and\ \bibinfo {author} {\bibfnamefont
  {R.}~\bibnamefont {Mattheis}},\ }\href {\doibase 10.1103/PhysRevB.65.020405}
  {\bibfield  {journal} {\bibinfo  {journal} {Phys. Rev. B}\ }\textbf {\bibinfo
  {volume} {65}},\ \bibinfo {pages} {020405} (\bibinfo {year}
  {2001})}\BibitemShut {NoStop}%
\bibitem [{\citenamefont {Rowan-Robinson}\ \emph {et~al.}(2017)\citenamefont
  {Rowan-Robinson}, \citenamefont {Stashkevich}, \citenamefont {Roussign{\'e}},
  \citenamefont {Belmeguenai}, \citenamefont {Ch{\'e}rif}, \citenamefont
  {Thiaville}, \citenamefont {Hase}, \citenamefont {Hindmarch},\ and\
  \citenamefont {Atkinson}}]{rowan2017interfacial}%
  \BibitemOpen
  \bibfield  {author} {\bibinfo {author} {\bibfnamefont {R.}~\bibnamefont
  {Rowan-Robinson}}, \bibinfo {author} {\bibfnamefont {A.}~\bibnamefont
  {Stashkevich}}, \bibinfo {author} {\bibfnamefont {Y.}~\bibnamefont
  {Roussign{\'e}}}, \bibinfo {author} {\bibfnamefont {M.}~\bibnamefont
  {Belmeguenai}}, \bibinfo {author} {\bibfnamefont {S.-M.}\ \bibnamefont
  {Ch{\'e}rif}}, \bibinfo {author} {\bibfnamefont {A.}~\bibnamefont
  {Thiaville}}, \bibinfo {author} {\bibfnamefont {T.}~\bibnamefont {Hase}},
  \bibinfo {author} {\bibfnamefont {A.}~\bibnamefont {Hindmarch}}, \ and\
  \bibinfo {author} {\bibfnamefont {D.}~\bibnamefont {Atkinson}},\ }\href
  {\doibase 10.1038/s41598-017-17137-z} {\bibfield  {journal} {\bibinfo
  {journal} {Sci. Rep.}\ }\textbf {\bibinfo {volume} {7}},\ \bibinfo {pages}
  {16835} (\bibinfo {year} {2017})}\BibitemShut {NoStop}%
\bibitem [{\citenamefont {Doh}\ \emph {et~al.}(2005)\citenamefont {Doh},
  \citenamefont {van Dam}, \citenamefont {Roest}, \citenamefont {Bakkers},
  \citenamefont {Kouwenhoven},\ and\ \citenamefont {De~Franceschi}}]{Doh272}%
  \BibitemOpen
  \bibfield  {author} {\bibinfo {author} {\bibfnamefont {Y.-J.}\ \bibnamefont
  {Doh}}, \bibinfo {author} {\bibfnamefont {J.~A.}\ \bibnamefont {van Dam}},
  \bibinfo {author} {\bibfnamefont {A.~L.}\ \bibnamefont {Roest}}, \bibinfo
  {author} {\bibfnamefont {E.~P. A.~M.}\ \bibnamefont {Bakkers}}, \bibinfo
  {author} {\bibfnamefont {L.~P.}\ \bibnamefont {Kouwenhoven}}, \ and\ \bibinfo
  {author} {\bibfnamefont {S.}~\bibnamefont {De~Franceschi}},\ }\href {\doibase
  10.1126/science.1113523} {\bibfield  {journal} {\bibinfo  {journal}
  {Science}\ }\textbf {\bibinfo {volume} {309}},\ \bibinfo {pages} {272}
  (\bibinfo {year} {2005})}\BibitemShut {NoStop}%
\bibitem [{\citenamefont {Deng}\ \emph {et~al.}(2012)\citenamefont {Deng},
  \citenamefont {Yu}, \citenamefont {Huang}, \citenamefont {Larsson},
  \citenamefont {Caroff},\ and\ \citenamefont {Xu}}]{doi:10.1021/nl303758w}%
  \BibitemOpen
  \bibfield  {author} {\bibinfo {author} {\bibfnamefont {M.~T.}\ \bibnamefont
  {Deng}}, \bibinfo {author} {\bibfnamefont {C.~L.}\ \bibnamefont {Yu}},
  \bibinfo {author} {\bibfnamefont {G.~Y.}\ \bibnamefont {Huang}}, \bibinfo
  {author} {\bibfnamefont {M.}~\bibnamefont {Larsson}}, \bibinfo {author}
  {\bibfnamefont {P.}~\bibnamefont {Caroff}}, \ and\ \bibinfo {author}
  {\bibfnamefont {H.~Q.}\ \bibnamefont {Xu}},\ }\href {\doibase
  10.1021/nl303758w} {\bibfield  {journal} {\bibinfo  {journal} {Nano Letters}\
  }\textbf {\bibinfo {volume} {12}},\ \bibinfo {pages} {6414} (\bibinfo {year}
  {2012})}\BibitemShut {NoStop}%
\bibitem [{Dat()}]{Dataset}%
  \BibitemOpen
  \href@noop {} {}\bibinfo {note} {{Nathan Satchell (2019): Supercurrent in
  ferromagnetic Josephson junctions with heavy metal interlayers II: canted
  magnetization - dataset University of Leeds. [Dataset].
  https://doi.org/10.5518/583}}\BibitemShut {NoStop}%
\end{thebibliography}%

\pagebreak

\onecolumngrid
\appendix

\begin{center}
\large{Supplementary Material for: ``Supercurrent in ferromagnetic Josephson junctions with heavy metal interlayers. II. Canted magnetization'' by Nathan Satchell, Reza Loloee, and Norman O. Birge}
\end{center}



\begin{figure}[h] 
\includegraphics[width=0.8\columnwidth]{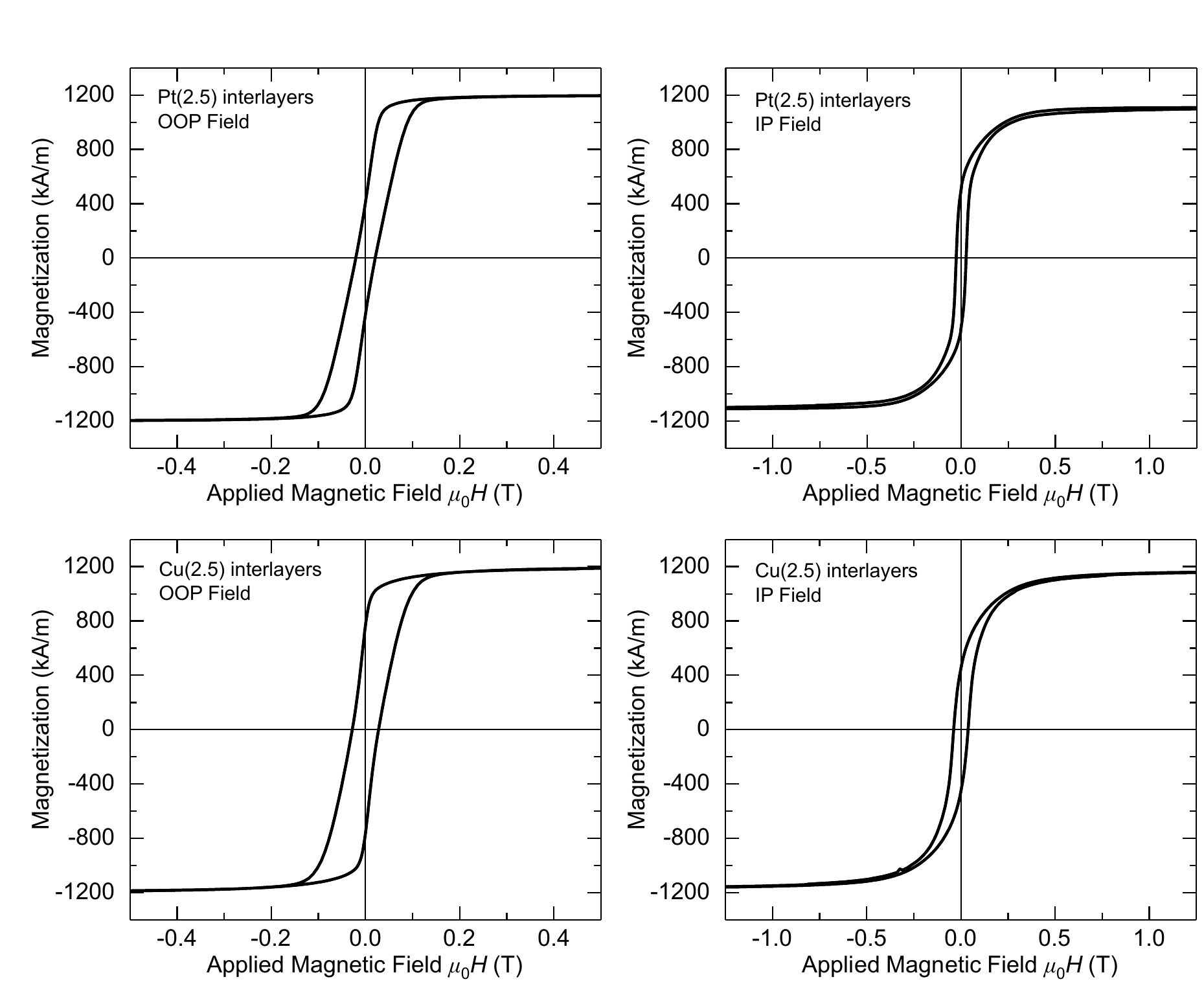} 
\caption{Magnetic hysteresis loops acquired at a temperature of 10~K for the series $S$--$N$--[Co(0.4)/Ni(0.4)]$_{\times 8}$/Co(0.4)--$N$--$S$ where the interlayer $N$ (either Pt(2.5) or Cu(2.5)) and applied field orientation (either in-plane (IP) or out-of-plane (OOP)) are indicated on the individual loops. The diamagnetic contribution from the substrate has been subtracted. Values of $M$ are calculated from the measured total magnetic moments and areas of the samples, and the total nominal thicknesses of the Co and Ni layers. The uncertainty in $M$ is dominated by the area measurements (different portions of the samples were used for each measurement), and is less than 5\%. (1~kA/m = 1~emu/cm$^3$).  \label{ARN}}
\end{figure}

\pagebreak

\begin{figure} [p]
\includegraphics[width=0.65\columnwidth]{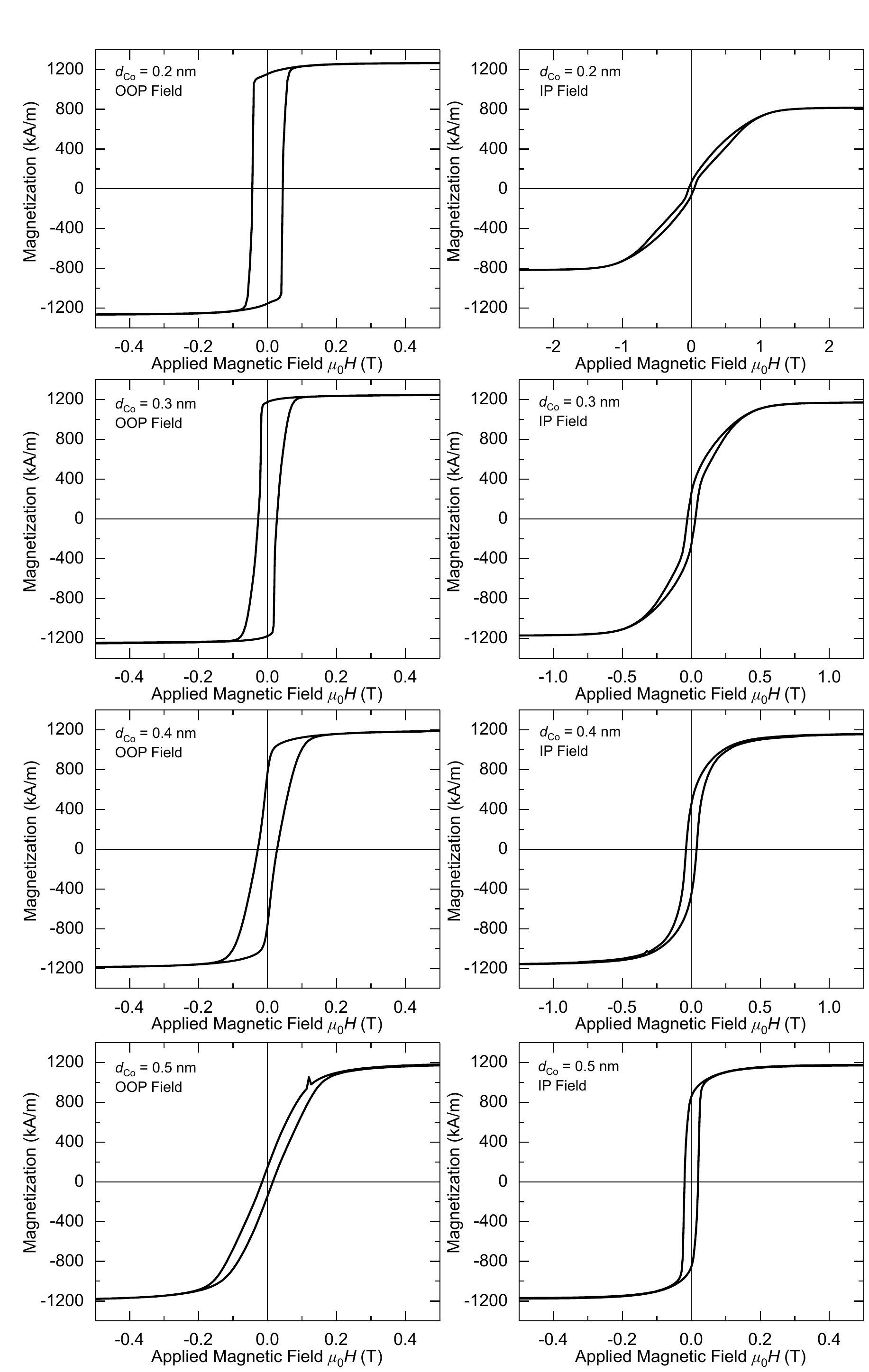} 
\caption{Magnetic hysteresis loops acquired at a temperature of 10~K for the series $S$--Cu(2.5)--[Co($d_\text{Co}$)/Ni(0.4)]$_{\times 8}$/Co($d_\text{Co}$)--Cu(2.5)--$S$ sheet film sample where the thickness of the Co and applied field orientation (either in-plane (IP) or out-of-plane (OOP)) are indicated on the individual loops. The diamagnetic contribution from the substrate has been subtracted. Values of $M$ are calculated from the measured total magnetic moments and areas of the samples, and the total nominal thicknesses of the Co and Ni layers. The uncertainty in $M$ is dominated by the area measurements (different portions of the samples were used for each measurement), and is less than 5\%. (1~kA/m = 1~emu/cm$^3$). \label{dCo}}
\end{figure}

\pagebreak

\begin{figure} [p]
\includegraphics[width=0.95\columnwidth]{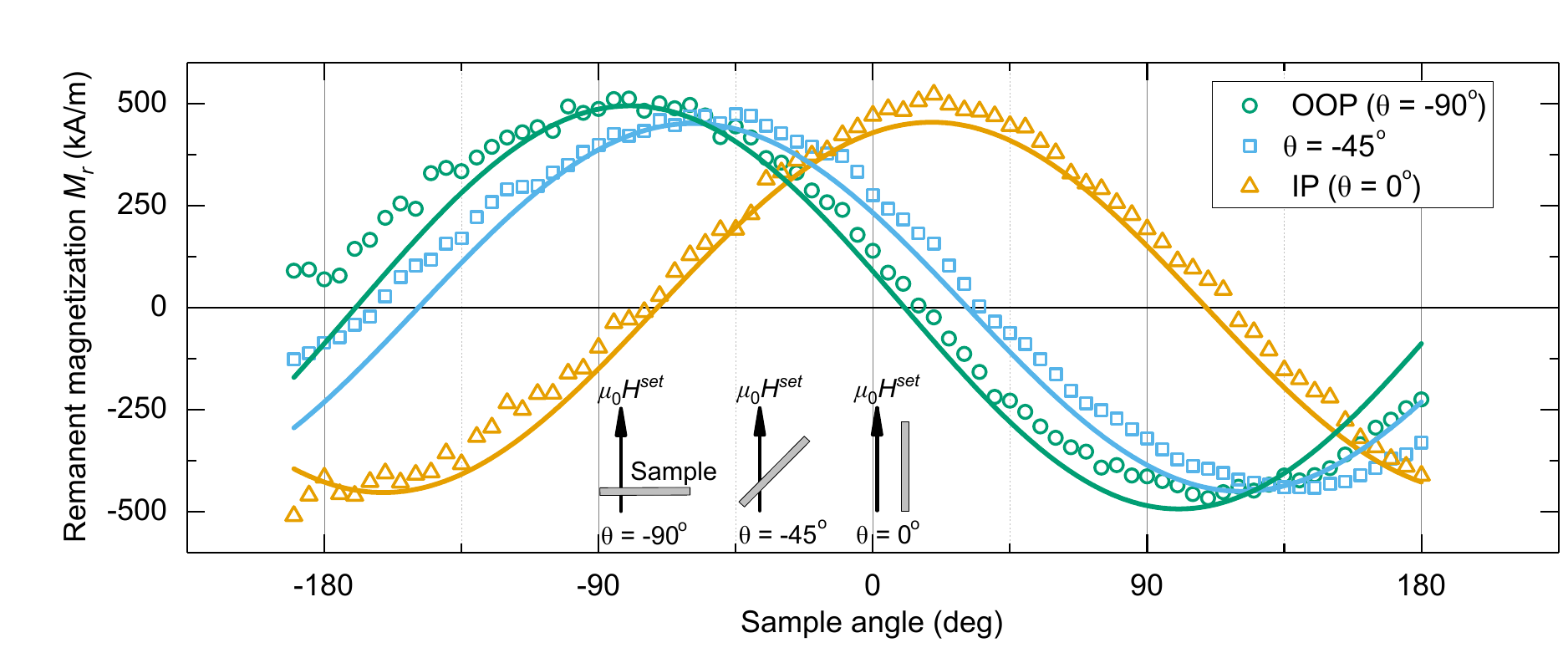} 
\caption{The zero field remanent magnetization at 10~K for the sample $S$--Pt(4.5)--[Co(0.4)/Ni(0.4)]$_{\times 8}$/Co(0.4)--Pt(4.5)--$S$ as a function of angle for a set field of 600~mT applied in the same orientations as Figure 1 (a-c) of the main text. The background signal due to the rotator has been subtracted. The lines are fits to a cosine function, $y=A\cos({x-x_c}$) where $A$ and $x_c$ are free fit parameters. The fits are suggestive of a canting away from the direction of applied field by $10^{\circ}\pm5^{\circ}$ from the OOP set field, $15^{\circ}\pm5 ^{\circ}$ from the $45^{\circ}$ set field, and $19^{\circ}\pm5^{\circ}$ from the IP set field. Due to the small portion of the sample used to fit onto the rotator holder and background signal from the rotator, the value of $M_r$ is less reliable than the data in the main text on the same sample, which was acquired using larger portions mounted in low background straws. The manufacturer quoted error in sample angle is 10$^{\circ}$, however our own characterization suggests reproducibility of the angle is better than 5$^{\circ}$, except at the extreme limits of sample travel. (1~kA/m = 1~emu/cm$^3$). \label{Fraunhofer}}
\end{figure}


\end{document}